\documentclass[12pt, preprint, linenumbers]{aastex631}
\let\splitbox\relax

\usepackage{adjustbox}
\usepackage{amsmath}
\usepackage{graphicx}
\usepackage{tabularray}
\usepackage{enumitem}

\usepackage{xcolor}

\begin{document}

\nolinenumbers
\newcommand{\argmin}{\mathop{\mathrm{argmin}}}  

\title{Red noise-based false alarm thresholds for astrophysical periodograms via Whittle's approximation to the likelihood}

\author{Amna Ejaz}
\affiliation{University of Delaware}
\email{amnaejaz@udel.edu}

\author{Sarah Dodson-Robinson}
\affiliation{University of Delaware}
\email{sdr@udel.edu}

\author{Charlotte Haley}
\affiliation{Argonne National Lab}
\email{haley@anl.gov}





\begin{abstract}
\nolinenumbers
Astronomers who search for periodic signals using Lomb-Scargle periodograms rely on false alarm level (FAL) estimates to identify statistically significant peaks. Although FALs are often calculated from white noise models, many astronomical time series suffer from red noise. Prewhitening is a statistical technique in which a continuum model is subtracted from log power spectrum estimate, after which the observer can proceed with a white-noise treatment. Here we present a prewhitening-based method of calculating frequency-dependent FALs. We fit power laws and autoregressive models of order 1 to each Lomb-Scargle periodogram by minimizing the Whittle approximation to the negative log-likelihood (NLL), then calculate FALs based on the best-fit model power spectrum. Our technique is a novel extension of the Whittle NLL to datasets with uneven time sampling. We demonstrate FAL calculations using observations of $\alpha$~Cen~B, GJ~581, HD 192310, synthetic data from the radial velocity (RV) Fitting Challenge, and {\it Kepler} observations of a differential rotator. The {\it Kepler} data analysis shows that only true rotation signals are detected by red-noise FALs, while white-noise FALs suggest all spurious peaks in the low-frequency range are significant. A high-frequency sinusoid injected into $\alpha$~Cen~B $\log R^{\prime}_{HK}$ observations exceeds the 1\% red-noise FAL despite having only 8.9\% of the power of the dominant rotation signal. In a periodogram of HD 192310 RVs, peaks associated with differential rotation and planets are detected against the 5\% red-noise FAL without iterative model fitting or subtraction. Software for calculating red noise-based FALs is available on GitHub.

\end{abstract}

\NewPageAfterKeywords


  \section{Introduction} \label{sec:intro}

Extreme-precision spectrographs with radial velocity precision approaching 10 cm~s$^{-1}$ are searching for Earth-mass planets orbiting sunlike stars~\citep{state,EXPRES,ESPRESSO}. Since the primary mathematical goal in an RV planet search is to detect and confirm a periodic signal in the time series, the typical first step is to examine a power spectrum estimate. The most commonly used power spectrum estimator for RV time series is the generalized Lomb-Scargle periodogram \citep[GLS periodogram;][]{zechmeister09}, though there are other period-search tools such as the Bayesian and stacked Bayesian GLS \citep{mortier2015bgls, mortier17}, the apodized Kepler periodogram \citep{gregory16}, compressed sensing techniques \citep{donoho2006compressed, nathanhara}, and Welch's method \citep{Welch, sdr}. 
When there is a peak in the periodogram, the peak frequency serves as the estimator for the frequency of the periodic process \citep[e.g.][]{montgomery99}. Possible sources of periodic signals are p-modes, planets, star rotation, and magnetic activity cycles. 

If the time series had infinite length, each periodogram peak would correspond to a true physical periodicity, but this is not the case in practice. Even if the spectrograph has extreme precision, 
the Lomb-Scargle periodogram \citep[as normalized by][]{scargle1982} gives spurious positive detections associated with statistical fluctuations above statistical significance levels given by $\chi^2_2$ quantiles \citep{ThomsonandHaley}. (See Appendix \ref{app:quantiles} for a discussion of the relationship between false alarm levels and $\chi^2_2$ quantiles).
To address the false positive problem and utilize error estimates in the observations, detection significance is quantified by the false alarm probability (FAP) that a time series with no true periodic signal could produce a periodogram peak with the observed power. The FAP calculation tools that are implemented in the \texttt{timeseries.LombScargle} class of the \texttt{astropy} software package \citep{astropy} are the analytical method of \citet{Baluev2008}, which approximates FAP as a function of periodogram power based on the theory of extreme values of stochastic processes~\citep{evt}, and the bootstrap method~\citep{Cumming, chernick2011bootstrap}, which involves creating an ensemble of noise series by randomly sampling the observations with replacement. Both methods assume that the time series contains an uncorrelated white noise background.

\begin{figure}
\centering
\includegraphics[width=\linewidth]{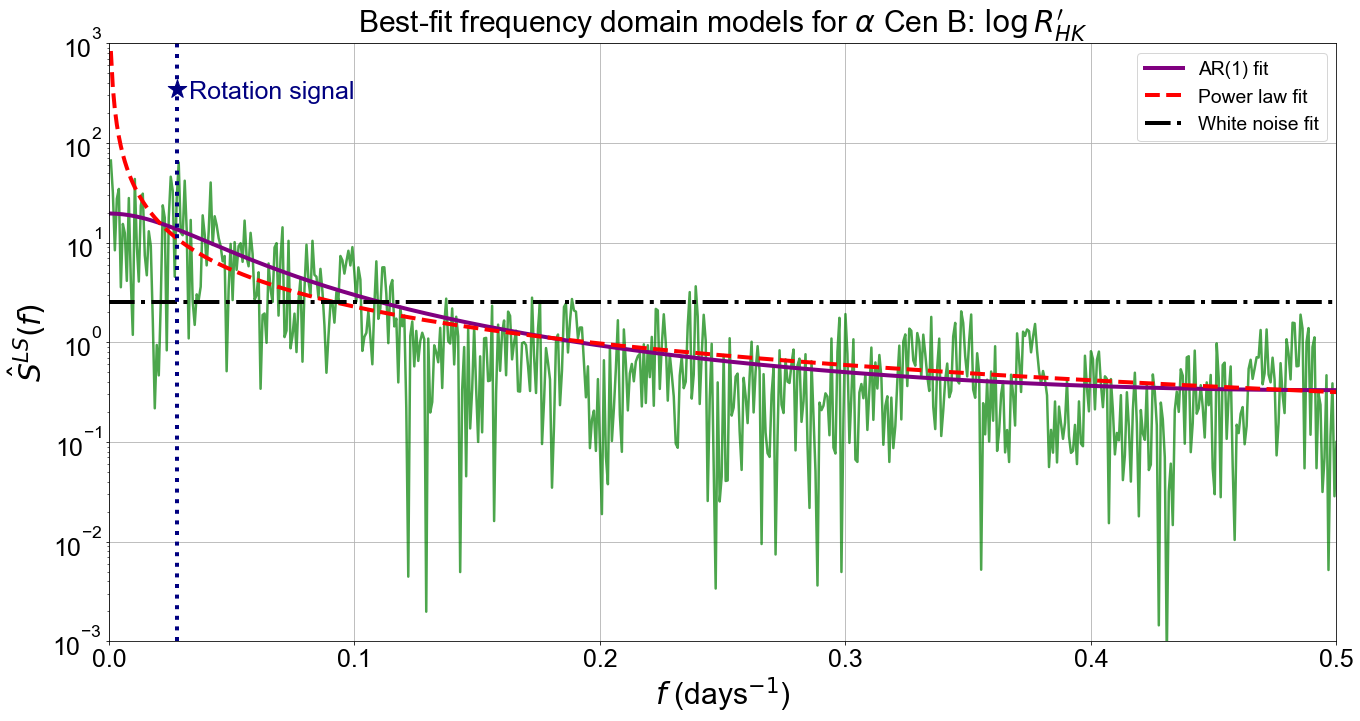}   
\caption{GLS periodogram $\hat{S}^{LS}(f_k)$ (green) of activity indicator $\log R^{\prime}_{HK}$ of $\alpha$~Cen~B~\citep{dumusque} with a red noise background. The vertical navy dotted line is the rotation signal, the horizontal black dash-dot line is a white noise fit to the spectrum, and the purple solid line and red dotted line are the best-fit versions of two red noise models, AR(1) and power law (see section~\S \ref{sec:models} for a detailed description of the red noise models used in this work).}
\label{fig:acenbfit}
\end{figure}

However, RV and magnetic activity-indicator time series often have a red noise background, which is indicated by a downward trend in a GLS periodogram plot with frequency on the x-axis and a logarithmic scale in the y-axis \citep[see Appendix B of][]{sdr}. For example, red noise is obvious in Figure~\ref{fig:acenbfit}, which shows the GLS periodogram of the $\alpha$~Cen~B $\log R_{HK}^{\prime}$ time series from \citet{dumusque}. The pioneering work by \citet{baluev2011} illustrates how red noise in the RVs of GJ~876 led to inaccurate estimates of orbital eccentricities and parameter uncertainties under the traditional white noise-based analysis method. Methods for computing red noise-based FAPs have been proposed by \citet{Vaughan2005} and \citet{Vaughan2010} for standard periodograms of time series with uniform timesteps and by \citet{Baluev2013, Baluev2015, Littlefair2017, LenoirCrucifix2018, delisle1, Hara2022} for GLS periodograms. These methods are based on models of the covariance matrix.

Just as noise properties can be inferred from a covariance matrix, they can also be estimated using parametric models in the frequency domain \citep{Kazlauskas2009TheCA, parametric_models}. For uniformly sampled data with constant timesteps, statisticians often {\it prewhiten} by fitting a continuum model to $\log [\hat{S}(f)]$ (where $\hat{S}(f)$ is the estimated power spectrum at frequency $f$), then subtracting the model \citep[e.g.]{prewhitening,Bayazit01082007}. The periodic signals are then identified as peaks in the residual spectrum at which the power exceeds a statistically significant $\chi^2$ threshold. In this work, we implement the first step of a prewhitening process by fitting a continuum model in the frequency domain. However, instead of subtracting the model from $\log [\hat{S}(f)]$ and searching for peaks in the residuals, we compute frequency-dependent false alarm levels that account for noise artifacts arising from uneven sampling, in which the timesteps vary.



Here we extend Whittle's negative log-likelihood approximation \citep[][see \S \ref{subsec:fitting} for a mathematical description]{whittle, original_whittle, whittle_2} for power spectrum model fitting to time series with uneven observing cadence. 
The best-fit models are used to calculate red noise-based false alarm levels (FALs), which are curves with constant FAP. 
Although the Whittle negative log-likelihood (NLL) has been applied to diverse topics such as hydrology \citep{whittle_hydrology}, econometrics \citep{whittle_econometrics}, and Earth's polar motion \citep{sykulski20}, this is the first use on unevenly sampled astronomical time series. 
Here we focus on planet-search data, but the method can be applied to any type of time series.

This paper is accompanied by a software package that implements Whittle NLL fitting of two classes of red-noise models: i) the autoregressive model of order 1 or AR(1) and ii) the power law. We explain the mathematical and physical motivations for each model in \S \ref{sec:models}. In \S \ref{sec:mm}, we show how to find the best-fit noise model using the Whittle NLL, then estimate false alarm levels.
We demonstrate the performance of our method on archival datasets in \S \ref{sec:analysis}, then discuss how our work complements other existing approaches in \S \ref{sec:comparison}. We present our conclusions and ideas for future work in \S \ref{sec:conclusion}. Appendix \ref{app:quantiles} examines how observational uncertainties affect FAL calculations. 


\section{Model Power Spectra} \label{sec:models}

Our software offers the user a choice between two red noise models. 
The first is the AR(1), a persistence model in which each observation is partially predicted by the immediately preceding one \citep{Percival,shumway}. 
The second is the frequency-domain power law, which is motivated by physical sources of red noise such as granulation and magnetic bright point motion~\citep{Kolmogorov,Frisch_1995, kippenhahn2012stellar,Cranmer, Cegla2018, Cegla2019,Paxton}. It is possible that multiple processes may contribute to the background noise, in which case more complex models like AR($p$) for $p > 1$; ARMA($p,q$) (autoregressive moving average); Mat\'{e}rn \citep[of which autoregressive models are a special case;][]{guttorp06}; and ARFIMA(p,d,q) \citep[autoregressive fractionally integrated moving average, of which power laws and autoregressive models are both special cases;][]{palma2007long} might describe the power spectrum better. Among the aforementioned models, ARFIMA is the most general. 


Here we have deliberately chosen simple models that are physically motivated and interpretable in order to introduce FAL calculation via the Whittle NLL. However, our methods work for any parametric power spectrum model.

\begin{figure}[h]
    \centering    
    \includegraphics[width=\linewidth]{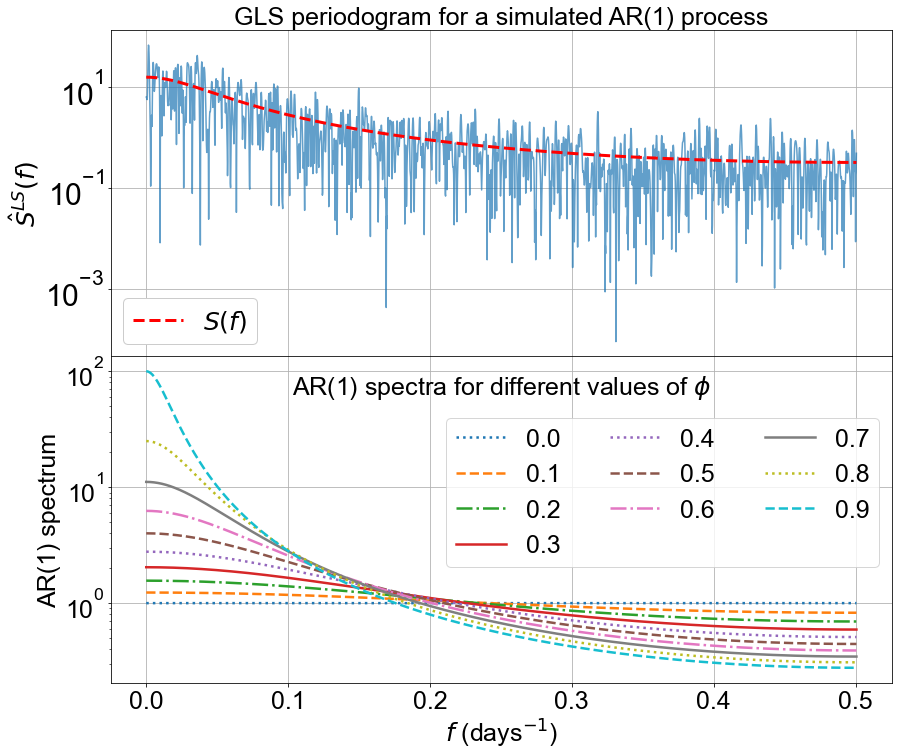}     
    \caption{{\bf Top} --- GLS periodogram $\hat{S}^{LS}(f_k)$~(blue) of a simulated AR(1) time series~(Eq.~\ref{eq:ar1_ts} with $\phi=0.75$ and $\sigma=1$). The dashed red line shows the analytic power spectrum $S(f)$~(Eq.~\ref{ar1ps}) of the AR(1), $\phi = 0.75$, $\sigma = 1$ process. {\bf Bottom} --- AR(1) power spectrum for different values of $\phi$.}
    \label{ar1-fig}
\end{figure}


\subsection{The AR(1) model: Autoregressive model of order 1}\label{sec:ar1}

In an autoregressive model of order p, or AR(p), each measurement is a linear combination of the $p$ previous observations plus a Gaussian noise term. A stationary AR(1) process sampled at uneven time intervals is defined as
\begin{align}\label{eq:ar1_ts}
    x_0 &\in \sigma_w \; \mathcal{N}(0, 1) \\
    x_{n} &= \phi^{(t_{n}-t_{n-1})/1} \; x_{n-1} + \zeta_{n} \\
    n &= 1, \ldots, N-1 
\end{align}
where $x_{n}$ is the measurement at time $t_{n}$ (note that $t_n - t_{n-1}$ is not constant), $t_n$ and $1$ have the same time units, $\mathcal{N}(0, 1)$ is the standard normal distribution, $N$ is the number of observations, and $\zeta_{n}$ is uncorrelated white noise with variance $\sigma_{w}^{2}$ \citep{Robinson, tauest}. The autoregressive coefficient $0 < |\phi| < 1$ is a measure of the correlation between successive observations; for red noise $0<\phi<1$. The larger the value of phi, the redder the periodogram, and the higher the ratio of the power at low frequencies to the power at high frequencies (Figure \ref{ar1-fig}, bottom panel). 
The persistence time $\tau$, which can be understood as the e-folding timescale over which observation $x_{n-1}$ has predictive power on $x_n$, is defined by
\begin{equation}\label{eqn:tau}
    \phi = \exp(-1 /\tau),
\end{equation}
where again $\tau$ and 1 have the same time units. 
The power spectrum of an AR(1) time series 
is
\begin{equation}\label{ar1ps}
    S(f) = \dfrac{\sigma^{2}_{w}}{|1-\phi \: e^{-2\pi if}|^{2}}
\end{equation}
\citep{shumway}, where $f$ is the frequency. The dashed red line in the top panel of Figure~\ref{ar1-fig} shows $S(f)$ for an AR(1) process 
with $\phi=0.75$ and $\sigma_w=1$, while the blue line shows the GLS periodogram $\hat{S}^{LS}(f_k)$ ($\hat{\cdot}$ denotes a statistical estimator, in this case of the power spectrum, and $f_k$ is is the $k^{\rm th}$ frequency gridpoint) of a simulated realization of the process.



\subsection{Power Law}\label{sec:pl}

Our software's second red-noise model option is the frequency-domain power law, which appears in many areas of stellar astrophysics. The predicted acoustic wave energy spectrum in the solar transition region follows a power law \citep{Cranmer}, as do the intensity variations of solar magnetic bright points \citep{cho19} and peak solar flare flux yields \citep{boffetta99}. Most notably, convective turbulence---in which energy from the largest eddies cascades down through smaller eddies until reaching the dissipation scale, where it is deposited as heat---is a source of red noise in time series of low-mass stars \citep[$M_* \leq 1.5 M_{\odot}$;][]{Frisch_1995, kippenhahn2012stellar, Cegla2018, Cegla2019, Paxton}.
The turbulent cascade covers an inertial range of eddy sizes over which we assume the energy flux remains constant and the energy transfer occurs with negligible viscous dissipation \citep{Chasnov,KATOPODES2019566}.

%

The inertial-range energy spectrum of a fully developed turbulent cascade, where all kinetic energy is transformed into heat at the dissipation scale, is described by the \citet{Kolmogorov} turbulence law,
\begin{equation}
    S(f) \propto f^{-5/3}.
\end{equation}
Kolmogorov's $5/3$ law assumes the turbulence is homogeneous (i.e.\ the flow statistics have no spatial dependence) and incompressible. 
However, these conditions may not be met in stellar atmospheres, leading to a different power-law exponent than -5/3 \citep{turbulence, turbulence2}.
Our power law model is
\begin{equation}\label{eqn:pl}
\log_{10} S(f) = -p \log_{10} f + a,
\end{equation}
which has parameter vector $\theta = \{ p, a\}$ with $p > 0$. 


\section{Mathematical and Computational Methods}\label{sec:mm}

Here we explain the mathematical and computational methods we use to fit red noise models and calculate FALs. We begin with an overview of the procedure, then describe steps \ref{step:bestfit}---\ref{step:FALs} in subsections \ref{subsec:fitting}---\ref{subsec:calculating}. We assume that the time series has had the sample mean subtracted, and prior to model fitting, we also standardize by dividing by the sample standard deviation so that the time series has unit variance, which ensures convergence for the AR(1) model fit.\footnote{The variance of an AR(1) process is $\sigma_{\rm AR(1)}^2 = \frac{\sigma_w^2}{1 - \phi^2}$. Since $0 < \phi < 1$ for red noise, standardizing such that $\sigma_{\rm AR(1)}^2 = 1$ ensures $\sigma_w^2$ is of order unity and an initial guess of $\sigma_w \lessapprox 1$ is close to the true solution.}



\begin{enumerate}

    \item \label{step:GLS} Estimate the generalized Lomb-Scargle periodogram~$\hat{S}^{LS}(f_k)$ on the frequency grid $f_k$ for $k = 0, 1, 2, \ldots, N_f - 1$ (where $N_f$ is the number of frequencies in the grid) using \texttt{astropy.timeseries.LombScargle}~\citep{astropy}.\footnote{The GLS periodogram is initially computed on a grid with a spacing of Rayleigh resolution $\mathcal{R}$, where $\mathcal{R} = 1/(t_{N-1} - t_0)$, for which the Whittle likelihood is defined. However, the calculation of FALs is done for an oversampled frequency grid with several samples per Rayleigh resolution.}
    To represent the power spectrum continuum shape as accurately as possible, we use the power spectral density (PSD) normalization option, which is comparable to the standard Schuster periodogram for even sampling and has $\mathrm{units} [\hat{S}^{LS}(f_k)] = \mathrm{units}[x^{2}_{n}] \cdot \mathrm{units}[t_n]$. Instrumental error bars are not used to calculate the PSD-normalized periodogram, though we use them to estimate FALs once the periodogram is computed (Step 3a). Weighting the $x_n$ by the error bars as described in \citet{zechmeister09} produces a dimensionless $\chi^2$ periodogram that discards information about the continuum shape, which is the defining characteristic of a red noise model.
    
    \item \label{step:bestfit} Find the best-fit AR(1), power law, and white noise model parameters for the GLS periodogram by minimizing the Whittle NLL \citep{whittle, Whittle1953TheAO, whittle1957curve}.\footnote{ To fit the AR(1) model, we convert the time units using $t^{\prime}_n = t_n / \operatorname{median}(\Delta t)$ so that the median timestep is 1 and the effective Nyquist frequency is $f_{\rm Nyq} = 0.5$~(median time units)$^{-1}$. This step improves the convergence of our fitting algorithm by ensuring that the distribution of $\phi^{(t_n - t_{n-1}) / 1}$ is centered at $\phi$ instead of a power of $\phi$. Before plotting the FALS, we convert back into the original time and frequency units.} Section \ref{subsec:fitting} describes the Whittle NLL and explains our motivation for extending it to astronomical time series.
    \item Select the model type that best describes the frequency-domain noise properties (\S \ref{subsec:assessing}):
    \begin{enumerate}
        \item Create a set of Monte Carlo realizations of the time series using published error bars, 
        \item Calculate the Whittle NLLs of the best-fit AR(1), power law, and white noise models for the GLS periodogram of each time series realization, 
        \item Choose the model type that tends to produce the lowest Whittle NLL. 
    \end{enumerate} 
    \item \label{step:FALs} Calculate false alarm levels~(FALs) based on the best-fit noise model (\S \ref{subsec:calculating}).
\end{enumerate}

We demonstrate steps \ref{step:bestfit}--\ref{step:FALs} using either the $\alpha$~Cen~B $\log R^{\prime}_{HK}$ time series of \citet{dumusque} or the GJ~581 H$\alpha$ index reported by \citet{GJ581-Robertson}.
$\log R_{HK}^{\prime}$ is a bolometric flux-corrected version of the Mt.\ Wilson S-index. H$\alpha$ and $\log R_{HK}^{\prime}$ are sensitive to energy release by magnetic reconnection in the stellar chromosphere \citep{vaughan78, logrhk}.


\subsection{Fitting the models}\label{subsec:fitting}

To estimate the parameter vector $\hat{\theta}$ of the best-fit model, we minimize the Whittle approximation to the Gaussian negative log-likelihood \citep[Whittle NLL;][]{whittle, Whittle1953TheAO, whittle1957curve}:
\begin{equation}\label{eq:wnll}
    -\mathcal{L}(\theta) \approx \sum\limits_{k=0}^{N_f-1}\bigg(\ln[S(f_k,\theta)]+\dfrac{\hat{S}^{LS}(f_k)}{S(f_k,\theta)}\bigg).
\end{equation}
In Equation \eqref{eq:wnll}, $-\mathcal{L}(\theta)$ is the Whittle NLL given by model power spectrum $S(f_k,\theta)$ evaluated at frequency gridpoints $f_k$, $\theta$ is the vector of model parameters, and $\hat{S}^{LS}(f_k)$ is the GLS periodogram.
The parameter vector estimate is
\begin{equation}
    \hat{\theta} = \argmin_{\theta} (-\mathcal{L}(\theta)).
    \label{eq:bestfit}
\end{equation}


\begin{figure}[h]
\begin{tabular}{c}
\includegraphics[width=0.8\textwidth]{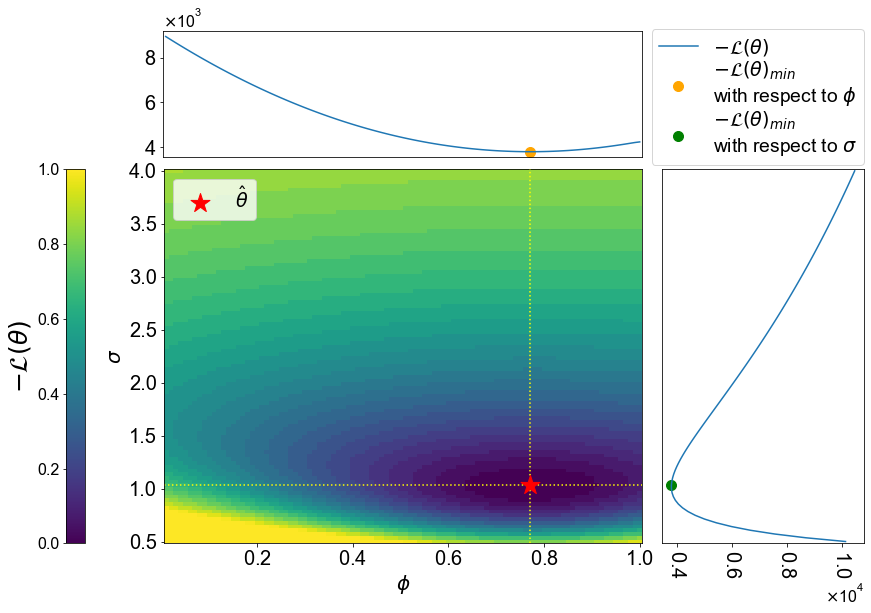} \\
\includegraphics[width=0.8\textwidth]{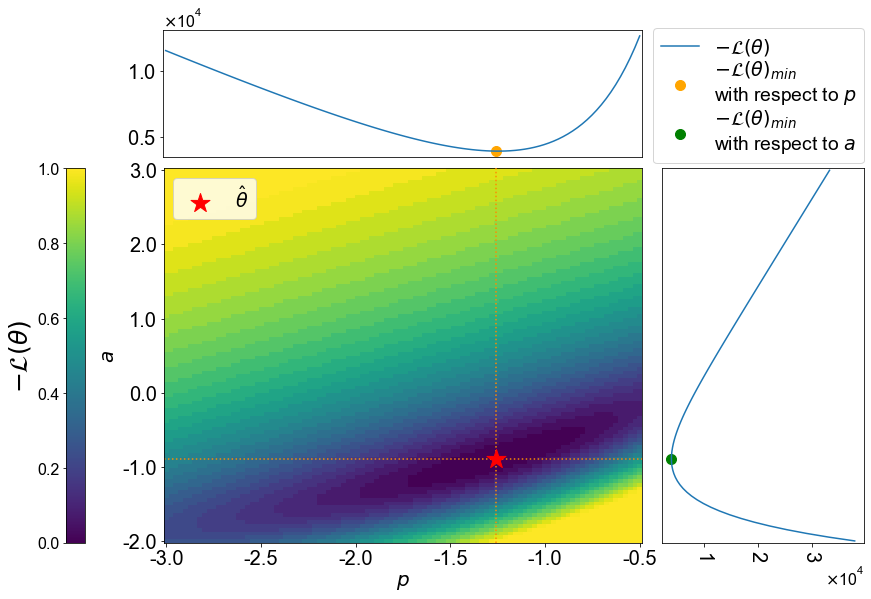} 
\end{tabular}
\caption{{\bf Top} --- A 2D color plot of the Whittle NLL function $-\mathcal{L}(\theta)$ against AR(1) model parameters for the GLS periodogram of the \citet{dumusque} $\alpha$~Cen~B $\log R^{\prime}_{HK}$ time series. The x-axis and y-axis of the colored plot represent $\phi$ and $\sigma_w$ respectively. {\bf Bottom} --- $-\mathcal{L}(\theta)$ for power law models of the same time series. Here the x-axis and y-axis represent normalization $a$ and exponent $p$ respectively. The color bars are normalized between 0 and 1 for visualization purposes. The blue curves show the marginalization of $-\mathcal{L}(\theta)$ with respect to a single parameter. The minimum values of $-\mathcal{L}(\hat{\theta})$ are marked with red stars.}
\label{fig:acenbmin}
\end{figure}


Figure \ref{fig:acenbmin} is a visualization of the Whittle likelihoods for the AR(1) and power-law model fits to the GLS periodogram of the $\alpha$ Cen B $\log R_{HK}^{\prime}$ time series from \citet{dumusque}. The top pseudocolor plot shows $-\mathcal{L}(\theta), \theta = \{\phi,\sigma_w\}$ for the AR(1) model, while the bottom pseudocolor plot depicts $-\mathcal{L}(\theta), \theta =\{p, a\}$ for the power law model. 
Each parameter space has a well-defined minimum $-\mathcal{L}(\hat{\theta})$ marked with a red star, though the power law model has more degeneracy between the two free parameters than the AR(1). The blue curves show marginalizations of the Whittle NLL over each parameter, with the best-fit value marked in orange.

We use the gradient-free Nelder-Mead method \citep{Nelder-mead} implemented in \texttt{scipy.optimize.minimize} to minimize the Whittle NLL~\citep{2020SciPy-NMeth}. For the sake of comparison, we also fit a white-noise model to each dataset:
\begin{equation}
    S(f) = c, 
\end{equation}
where $2c/f_{\rm Nyq}$ (where $f_{\rm Nyq}$ is the maximum frequency in the grid, or pseudo-Nyquist frequency) is the variance of the white noise and $\hat{\theta}_{wn}=\{\hat{c}\}$. Note that white noise is a special case of both AR(1) (with $\phi =0$) and power law (with $p=0$) models. Figure~\ref{fig:acenbfit} shows the best-fit AR(1) (purple solid line), power law (red dashed line), and white noise (blue dash-dot) model fits to the GLS periodogram of the $\alpha$~Cen~B $\log R_{HK}^{\prime}$ time series \citep{dumusque}.

Investigators in fields such as statistics~\citep{HAUSER1999229}, physiology~\citep{Roume2023}, and psychology~\citep{Altamore_undated-nc} have used the Whittle NLL to optimize the ARFIMA(p,d,q) model, which is a general model class that encompasses power laws, autoregressives, and moving averages. ARFIMA models are also occasionally used in astronomy \citep{2009ApJ...693.1877S,2019MNRAS.485.3970S,2018FrP.....6...80F,2019AJ....158...58C}. R/CRAN packages for fitting ARFIMA models using the Whittle likelihood include \texttt{beyondWhittle}~\citep{beyondWhittle,beyondWhittle1,beyondWhittle3} and \texttt{LSTS}~\citep{LSTS}. However, so far the ARFIMA formalism has only been applied to power spectrum estimates computed from uniformly sampled time series. Calculation of FALs for GLS periodograms using ARFIMA-based noise models is a natural extension of this work and can be built on top of our computational framework.

\subsection{Model selection}
\label{subsec:assessing}

\begin{figure}[h]
    \centering
    \includegraphics[width=0.8\textwidth]{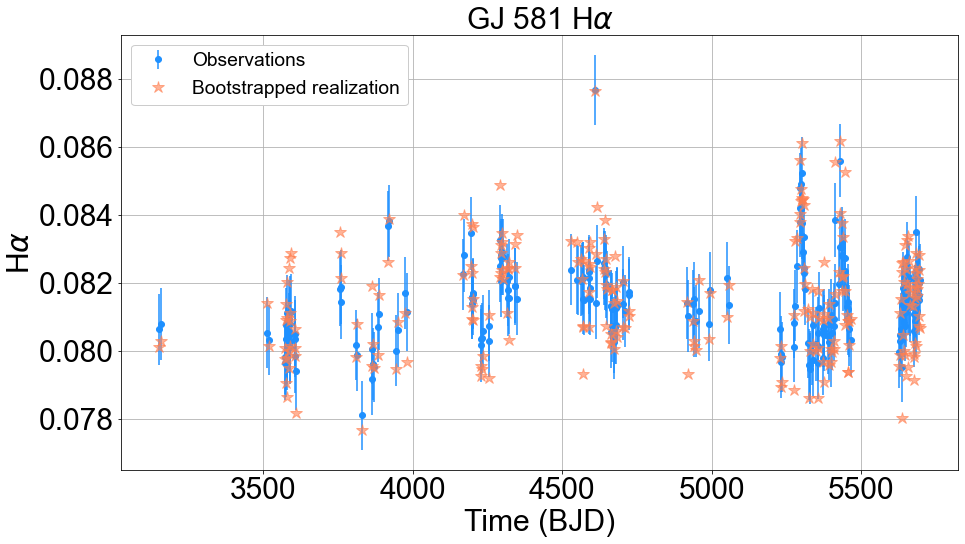}
    \caption{The observed H$\alpha$ time series of GJ 581~\citep[][blue]{GJ581-Robertson} and a new realization created by adding white noise to each observation (red).}
    \label{fig:newts}
\end{figure}

With optimized parameters of the two power spectrum model types in hand, the next step is to choose the model type that best describes the dataset. While we could simply pick the model that gives the lowest value of $-\mathcal{L}(\theta)$, we opt for a more robust comparison. Ten thousand new time series $x^{\star}_n$ are generated by adding independent white noise with mean zero and standard deviation given by the error bar at each time stamp, such that $x^{\star}_n = x_n + \mathcal{N}(0, \sigma_n)$. (Error bars are assumed to be temporally uncorrelated,  though our methods could be extended to treat correlated error bars using moving average models.)
An example using the H$\alpha$ time series of GJ 581 \citep{GJ581-Robertson} is shown in Figure \ref{fig:newts}. 
We then fit the AR(1), power law, and white noise models to the GLS periodogram of each realization. This experiment creates a distribution of Whittle NLL values for each model type, which we can compare before committing to a particular model.


\begin{figure}[h]
\centering
\includegraphics[width =0.95\linewidth]{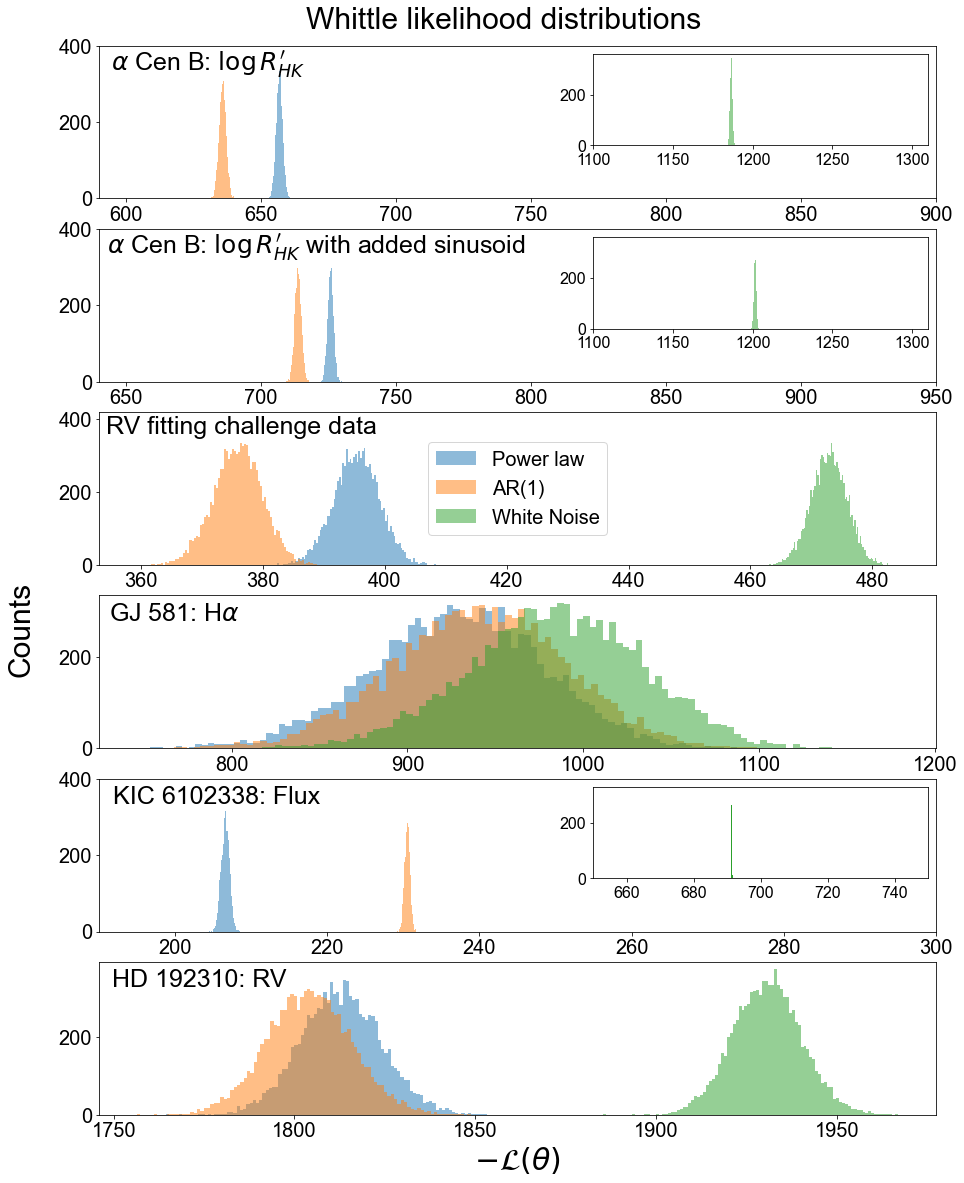}
\caption{The Whittle NLL distributions created by fitting all three noise models to GLS periodograms of 10000 realizations of each time series studied in this paper. $-\mathcal{L}(\theta)$ distributions from the AR(1) model are colored orange, $-\mathcal{L}(\theta)$ distributions from the power-law model are in blue, and  $-\mathcal{L}(\theta)$ distributions from the white-noise fits are in green. From top to bottom, the time series are $\alpha$~Cen~B $\log R^{\prime}_{HK}$ \citep{dumusque}, simulated RVs from System 11 of the RV Fitting Challenge \citep{RVchallengedata}, GJ~581 H$\alpha$ \citep{GJ581-Robertson}, {\it Kepler} photometry of KIC~6102338 \citep{borucki10} and HD 192310 RVs \citep{Laliotisetal2023AJ....165..176L}.}
\label{fig:wnll}
\end{figure}



The top panel of Figure \ref{fig:wnll} shows the $-\mathcal{L}(\theta)$ distributions that result from fitting each model type to the GLS periodogram shown in Figure \ref{fig:acenbfit}, which comes from the $\alpha$~Cen~B $\log R^{\prime}_{HK}$ time series of \citet{dumusque}. From Figure \ref{fig:wnll}, we see that the white noise model produces much higher values of $-\mathcal{L}(\theta)$ than either red noise model (see inset). The Whittle NLL distribution positioned at the lowest values of $-\mathcal{L}(\theta)$ belongs to the AR(1) model, indicating that AR(1) is a better choice than a power law for computing false alarm levels.



\subsection{Calculating the false alarm levels}
\label{subsec:calculating}

After choosing a model for the GLS periodogram, the observer can use time-domain realizations of the model to calculate false alarm levels. Model realizations are created in the following manner:

\begin{itemize}
    \item \textbf{AR(1):} After first selecting $x_0(t_0)$ from a normal distribution, use Equation~\eqref{eq:ar1_ts} recursively to obtain $x_n(t_n)$.
    \item \textbf{Power law:} 
    The time-domain form of a stochastic process with a power-law spectrum is \citep{powerlawts}
    \begin{equation}
        x_n = \sum_{k = 0}^{N_{f}-1} \sqrt{S(f_k,\hat{\theta}_{PL}) 2\pi \Delta f} \cos{(2\pi f_k t_{n}+\varphi_{k})}.
        \label{eq:timedomain_powerlaw}
    \end{equation}
    In Equation \eqref{eq:timedomain_powerlaw}, $S(f_k,\hat{\theta_{PL}})$ is the model power spectrum sampled on the frequency grid (Eq~\ref{eqn:pl}) and $\varphi_{k} \in [0, 2\pi]$ is the randomly chosen phase at each frequency, which is the source of fluctuations in the time series.
\end{itemize}

We generate 10,000 realizations of the chosen model with parameters randomly selected from the $\hat{\theta}$ distribution produced in the Monte Carlo procedure in \S \ref{subsec:assessing}. We sample from the posterior distribution of $\hat{\theta}$ instead of the single best-fit parameter vector so that any correlations between the parameters manifest in the FALs. 
From the GLS periodograms $\hat{S}^{LS}(f_k)$ of the 10,000 time-domain noise model realizations, we use the 95\%, 99\%, and 99.9\% percentile at each frequency to define the 5\%, 1\%, and 0.1\% false alarm levels.


\begin{figure}[h]
\centering
\includegraphics[width=\linewidth]{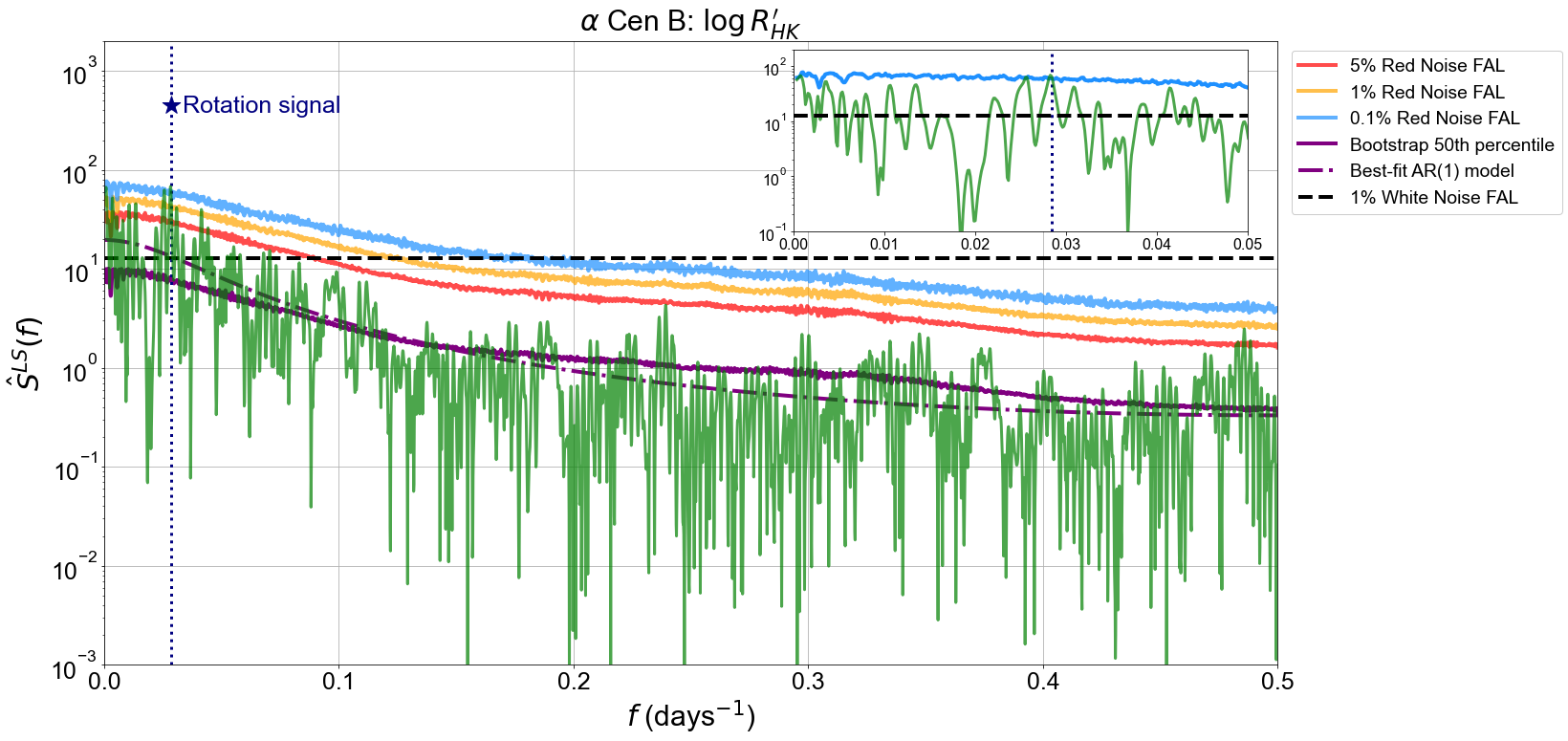}

\caption{AR(1)-based false alarm levels for the GLS periodogram of the $\alpha$ Cen B $\log R_{HK}^{\prime}$ time series \citep{dumusque} The $5\%$~(red line), $1\%$~(orange), and $0.1\%$~(blue) red-noise FALs follow the shape of the red-noise continuum. The $1\%$ white noise FAP estimated by the bootstrap method is plotted for comparison (black dashed line). The best-fit AR(1) model~(purple dash-dot line) and the $50\%$ percentile~(purple line) of Monte Carlo realizations are also shown. The rotation signal~(navy dotted line) reaches the 0.1\% FAL or $99.9\%$ significance threshold. The figure inset at the top right shows a zoomed-in version of the plot centered on the rotation signal (navy dotted line), which reaches the 0.1\% FAL.}
\label{fig:acenb_faps}
\end{figure}


Figure \ref{fig:acenb_faps} shows the FALs computed from the AR(1) model applied to the $\alpha$~Cen~B $\log R^{\prime}_{HK}$ periodogram. The AR(1)-based FALs alleviate one of the main problems that come from using a white noise model on GLS periodograms with a red noise background---the tendency to over-predict the significance of low-frequency signals. While $\hat{S}^{LS}(f_k)$ exceeds the white noise-based 1\% FAL at most frequencies $f < 0.06$~day$^{-1}$, the only signals that exceed the AR(1)-based 1\% FAL are at the known rotation frequency, $f = 1/38$~day$^{-1}$ \citep[][navy dotted line in Figure \ref{fig:acenb_faps}]{dumusque}, and at $f = 0.042$~day$^{-1}$ ($P = 23.98$~days).

As a diagnostic of how closely the Monte Carlo realizations of the power spectrum align with the best-fit model power spectrum, Figure \ref{fig:acenb_faps} shows the best-fit model (purple dash-dot line) and the 50\% percentile of the Monte Carlo realizations (purple solid line). For a dataset with negligible error bars and a perfectly fitted model, the two curves would match. Mismatches come from two sources: first, the continuum of the GLS periodogram might depart from the AR(1) or power law framework. Such departures are likely to occur with planet-search data, which have multiple noise sources that are not fully understood \citep[e.g.][]{palumbo24}. The bends in the 50\% percentile Monte Carlo curve at $0.14$~day$^{-1}$ and $0.33$~day$^{-1}$ probably represent such a mismatch. Serious differences between the two curves are a sign that a different (probably more complex) model power spectrum is required. The second source of mismatches between the model power spectrum and the 50\% percentile Monte Carlo curve is the tendency of error bars---which are treated as white noise terms in this analysis---to ``whiten'' the Monte Carlo realizations so that the dynamic range of the FALs is slightly lower than what the best-fit model would predict. In Appendix \ref{app:quantiles}, we use $\chi^2_2$ quantiles to demonstrate the effects of error bars on the FALs.



\section{Examples of red-noise based false alarm levels using archival data}
\label{sec:analysis}


\begin{table}
    \centering
    \noindent%
\begin{tblr}{width=\linewidth,hlines,vlines,cells={halign=c},cell{3}{4} = {yellow!40},cell{4}{4} = {yellow!40},cell{5}{4} = {yellow!40},cell{6}{6} = {yellow!40},cell{7}{6} = {yellow!40},cell{8}{4} = {yellow!40}}

&\SetCell[c=3]{c}{AR(1)}
     &  &  &  \SetCell[c=2]{c}{Power law }
        &  &   \SetCell[c=2]{c}{White noise}\\
     & $\theta_{AR(1)}$ & $\tau$ (d) & -$\mathcal{L}(\theta)_{min}$& $\theta_{PL}$ & -$\mathcal{L}(\theta)_{min}$& $\theta_{WN}$ & -$\mathcal{L}(\theta)_{min}$\\
     $\alpha$~Cen~B & $\{0.77,1.02\}$ & $3.83$ & $635.65$& $\{-1.23,-0.87\}$ & $656.70$& $\{2.55\}$ & $1186.70$\\
     $\alpha$~Cen~B + sinusoid & $\{0.74,1.09\}$ & $3.32$ & $713.61$& $\{-1.15,-0.76\}$ & $725.68$& $\{2.6\}$ & $1201.49$\\
     RV Fitting challenge & $\{0.84,0.7\}$ & $5.73$ &$373.83$& $\{-1.22,-1.02\}$ & $393.56$& $\{4.79\}$ & $472.11$\\
     GJ~581~H${\alpha}$ & $\{0.46,1.02\}$ & $2.45$ & $656.85$& $\{-0.64-0.44\}$ & $643.07$& $\{1.54\}$ & $765.59$\\
     KIC~6102338 & $\{0.99,0.14\}$ & $1.98$ & $230.48$& $\{-2.54,-4.27\}$ & $206.54$& $\{9.74\}$ & $691.35$\\
     HD 192310 & $\{0.81,1.16\}$ & $4.74$ & $1797.89$& $\{-0.73,0.05\}$ & $1807.16$& $\{15.51\}$ & $1926.88$\\  
\end{tblr}
\caption{The values of the best-fit parameters and the minimum Whittle NLL for each noise model for all time series studied in this paper. The shaded yellow cell in each row represents the lowest value of $-\mathcal{L}$ among the three noise models.}
\label{tab:tab1}
\end{table}


Here we demonstrate our method for calculating red noise-based false alarm levels using five different datasets: (i) the $\alpha$~Cen~B $\log R_{HK}^{\prime}$ time series with an injected high-frequency sinusoid \citep[][\S \ref{subsec:acenb}]{dumusque}; (ii) a simulated planet-search dataset created for the RV Fitting Challenge by~\citet{RVchallengedata}; (iii) the GJ~581 H$\alpha$ time series from~\citet{GJ581-Robertson}; (iv) {\it Kepler} observations of the differential rotator KIC 610233~\citep{borucki10}; (v) HD 192310 radial velocities~\citep{Laliotisetal2023AJ....165..176L}.

\subsection{$\alpha$ Cen B $\log R_{HK}^{\prime}$ with injected sinusoid}
\label{subsec:acenb}

\begin{figure}[h]
    \centering
    \includegraphics[width=0.8\linewidth]{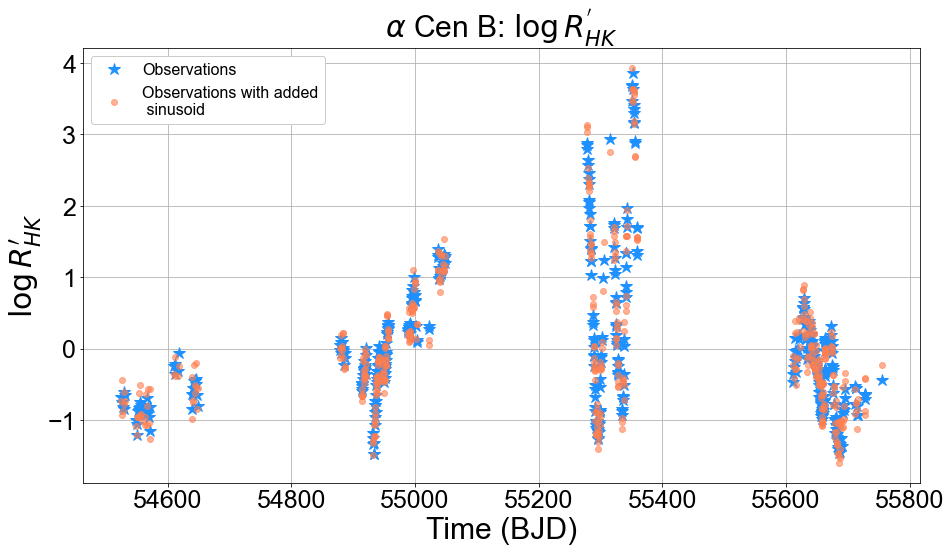}
    \caption{$\alpha$~Cen~B $\log R_{HK}^{\prime}$ data from \citet{dumusque}~(blue stars) and the same data set with an added periodic signal~(orange dots) with a period of $3$ days~(Equation~\ref{eqn:acenbsin}). 
    }
    \label{fig:acenb+sin}
\end{figure}

In \S \ref{subsec:calculating}, we discussed how the low-frequency continuum of a power spectrum with red noise can spuriously exceed a white noise-based FAL. Another problem with white noise-based FALs is that genuine high-frequency periodic signals may not appear to be statistically significant because they don't sit on top of a high continuum. Our goal is to prewhiten the power spectrum estimate and search for periodic signals with power that significantly exceeds the continuum. We demonstrate our method's performance on high-frequency signals by injecting a sinusoid into the \citet{dumusque} $\alpha$~Cen~B $\log R^{\prime}_{HK}$ time series:
\begin{equation}
    x_{n}^{\prime} = x_{n} + 0.25\sin(2\pi f_{0}t_{n}),
    \label{eqn:acenbsin}
\end{equation}
where $f_{0}=1/3$ day$^{-1}$. 
Error bars on $x_{n}^{\prime}$ match those of $x_n$. Figure \ref{fig:acenb+sin} shows $x_n$ (blue stars) and $x_{n}^{\prime}$ (orange dots).



\begin{figure}[h]
    \centering
    \includegraphics[width=\linewidth]{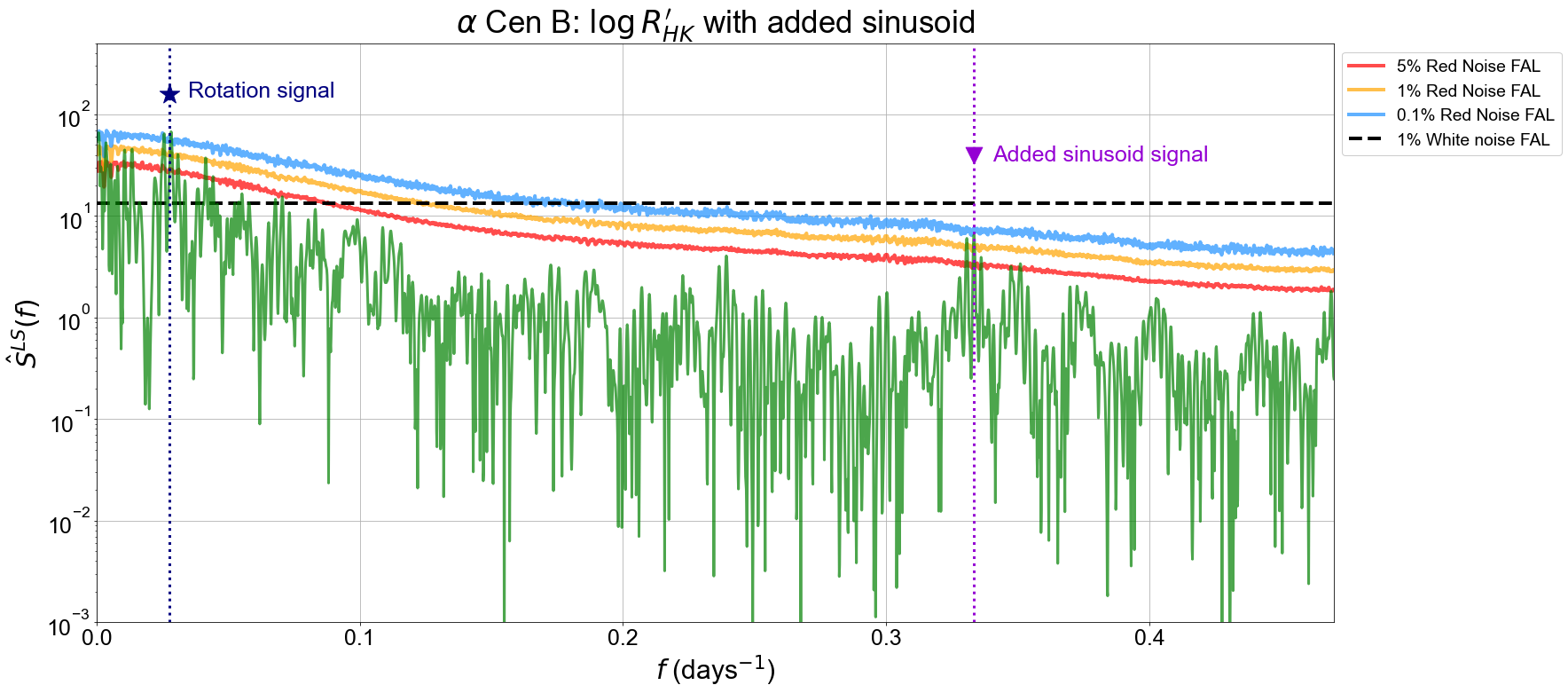}
    \caption{The GLS periodogram of the new time series $x_{n}^{\prime}$ generated by adding a periodic signal to the $\alpha$~Cen~B $\log R_{HK}^{\prime}$ data~\citep{dumusque} according to Equation~\ref{eqn:acenbsin}. The navy dotted line represents the rotation signal, and the light purple dotted line shows the added sinusoid. The added signal exceeds the 1\% FAL / $99\%$ significance level from the AR(1) model (blue solid line), but misses the white noise-based 1\% FAL (black dashed line).}
    \label{fig:acenb+sin_faps}
\end{figure}


Figure \ref{fig:wnll} (second row) depicts the $\mathcal{L}(\hat{\theta})$ distributions resulting from fitting the AR(1), power law, and white noise models to 10,000 realizations of $x_{n}^{\prime}$, while Figure \ref{fig:acenb+sin_faps} overlays the 1\% FAL computed from the AR(1) model (blue line) and the bootstrap 1\% white-noise FAL (red dashed line) on the GLS periodogram of $x_{n}^{\prime}$. Despite the fact that the estimated power at $f_{0}=1/3$ day$^{-1}$ is less than $10$\% of the power at the rotation frequency, we recover the added sinusoid with FAP$< 1$\% ($99\%$ detection significance). Since a white-noise model cannot accurately describe a power spectrum estimate that has not been prewhitened, the added sinusoid does not reach the $1\%$ white noise-based FAL (nor the 5\% white noise-based FAL, which is not shown). This experiment illustrates the importance of incorporating correlated noise into the calculation of FALs.


\subsection{RV Challenge dataset}
\label{subsec:rvchallenge}

\begin{figure}[h]
    \centering
    \includegraphics[width=0.8\linewidth]{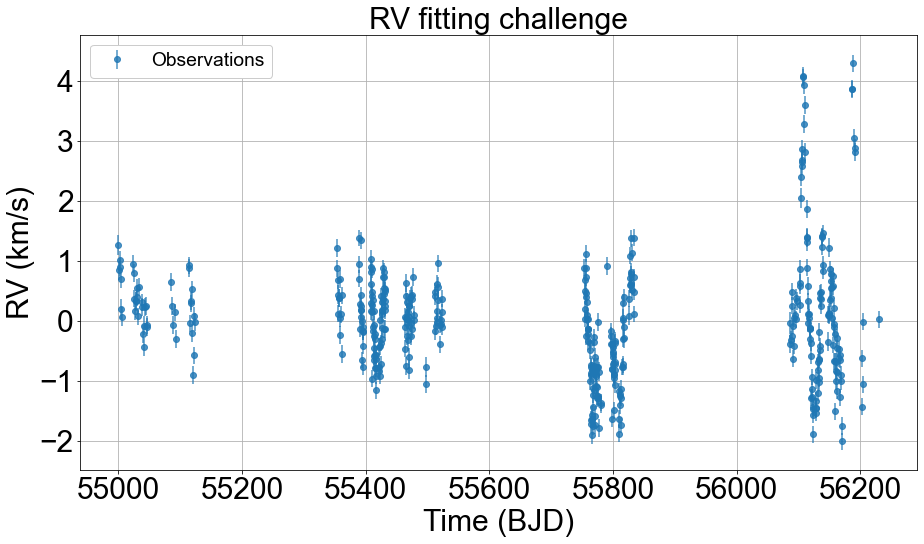}
    \caption{The simulated RV data set generated by \citet{RVchallengedata} for the 2016 RV Fitting Challenge~\citep{fitting}.}
    \label{fig:RV}
\end{figure}

Next, we compute red noise-based FALs for the GLS periodogram of a simulated RV data set generated by \citet{RVchallengedata} for the 2016 RV Fitting Challenge~\citep{fitting} (Figure \ref{fig:RV}). The challenge goal was to compare the performance of different methods for identifying the RV signals of low-mass planets in time series contaminated by stellar noise. The datasets were created by injecting Keplerian RV signals into stellar activity simulated by the SOAP~2.0 code \citep{soap20}, which models convective inhibition, plages, and spots. For our demonstration, we use System 11, which consists of the real RV measurements of $\alpha$~Cen~B from ~\citet{dumusque} plus five injected Keplerians. Three of the Keplerians correspond to the reported planets $\tau$~Ceti (HD~10700) b, c, and d, with periods $14.66$~days, $34.65$~days, and $96.93$~days, respectively \citep{Toumi2013_SignalstauCeti}. The remaining two signals are dummy planets added with periods of $283.10$ days and $3245.20$ days. In addition to the injected planets, the dataset inherited the $\alpha$~Cen~B rotation signal with a $\sim 37$-day period.


\begin{figure}[h]
    \centering   
    \includegraphics[width=\linewidth]{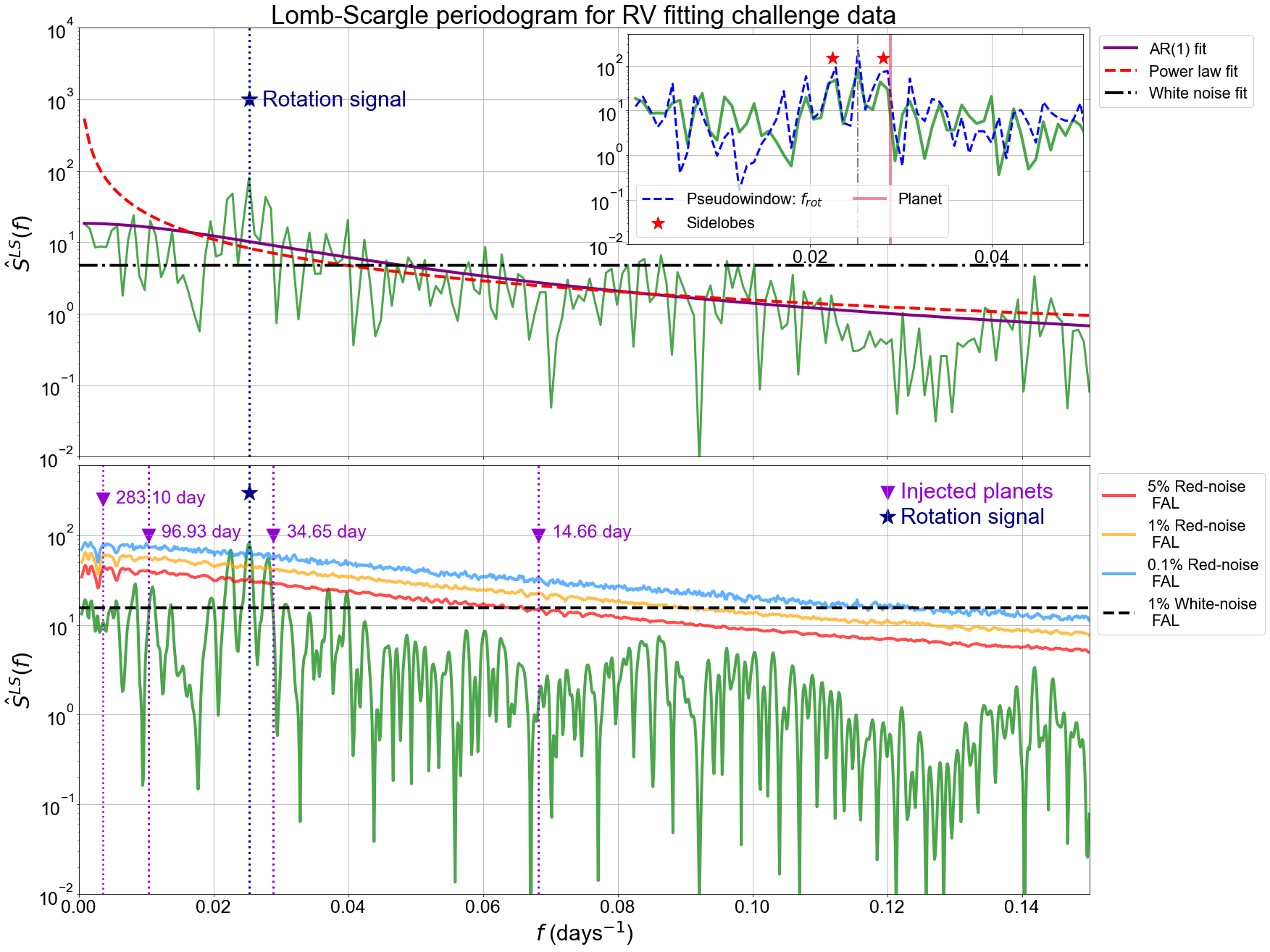}   
    \caption{RV fitting challenge data set~\citep{RVchallengedata}: {\bf Top}---The GLS periodogram $\hat{S}^{LS}(f_k)$~(green) with the best-fit AR(1)~(purple line), power law~(red dashed line) and white noise (blue dash-dot line) model fits. The navy dotted lines show the rotation signal. The inset plot shows the pseudowindow at the rotation frequency~(dark blue dashed line). From the pseudowindow plot, we see how the injected planet with period 34.65 days (pink vertical line) lands right on top of one of the side lobes (red star) of the rotation signal. {\bf Bottom}---The red-noise $5\%$~(red), $1\%$~(orange), $0.1\%$~(blue) FALs, and the $1\%$ white-noise FAL (black dashed line) are shown. The vertical dotted light purple lines, along with the marked periods (light purple triangles), represent the injected planets. Only the rotation signal~(navy dash-dot line) reaches a $0.1\%$ red-noise FAL.}
    \label{fig:RVchallengedata}
\end{figure}

Figure \ref{fig:RVchallengedata}~(top) shows the GLS periodogram $\hat{S}^{LS}(f_k)$ (green) with its best-fit AR(1) (purple line), power law (red dashed line) and white noise (black dash-dot line) models, calculated as in \S \ref{subsec:fitting}. The rotation signal and its first harmonic are marked with navy blue dotted lines. The best-fit model parameters $\hat{\theta}$ and the Whittle NLL $-\mathcal{L}(\hat{\theta})$ for all three models are shown in Table~\ref{tab:tab1}. Examining Figure~\ref{fig:wnll} (third row), we see that the AR(1) is the best-fit model for the GLS periodogram, as it tends to produce the lowest values of $-\mathcal{L}(\hat{\theta})$. 
The white noise model produces significantly higher Whittle NLL values than both red noise models.


Figure~\ref{fig:RVchallengedata}~(bottom) shows the $5\%$~(red), $1\%$~(orange), and $0.1\%$~(blue) false alarm levels based on the AR(1) model. The rotation signal~(navy dotted line), which is the most powerful signal in the data, reaches a $99.9\%$ significance threshold (0.1\% FAL). However, none of the injected planetary signals reach even the 5\% FAL. The dummy planet with $P = 3407.46$~days would be undetectable no matter the observing cadence, type of power spectrum estimator, or FAL computation procedure because its period is longer than the time baseline of the observations \citep[e.g.][]{godin1972analysis, loumos78, ramirezdelgado24}. The HD~10700~b and HD~10700~d clones and the dummy planet with $P = 283.10$~days all have semi-amplitudes of $K < 0.6$~m~s$^{-1}$. These signals were not detected by any team in the RV fitting challenge~\citep{fitting} and have amplitudes lower than the HARPS pre-2014 instrumental jitter, which is treated as a white-noise term \citep[e.g.][]{pepe11}.


The final injected planet, the HD~10700~c clone, has $K = 0.62$~m~s$^{-1}$ and $P = 34.65$~days ($f=0.0289$~day~$^{-1}$). To assess the detectability of this planet, we examine the inset plot in the top panel of Figure~\ref{fig:RVchallengedata}, which shows the {\it pseudowindow} for the rotation frequency ($f_{\rm rot} = 0.0270$~day$^{-1}$) overlaid on the GLS periodogram of the data. The pseudowindow at $f_0$ is the frequency response to the signal $x_n = \cos (2 \pi f_0 t_n)$ \citep{scargle1982}. By definition, any power in the GLS periodogram at frequencies other than $f_0$ is spurious. The pseudowindow plot shows that a sinusoid with $f = f_{\rm rot}$ yields a complex periodogram peak with sidelobes at $f_{\rm rot} \pm 1/365.25$~day$^{-1}$ and $f_{\rm rot} \pm 2/365.25$~day$^{-1}$ caused by the yearly alias. The injected planet has a frequency that places it within one of the rotation signal sidelobes, making it impossible to detect from the RV periodogram alone---a mathematical method that disentangles stellar and planetary signals based on activity indicators and/or CCFs is required \citep[e.g.][]{EXPRES-challenge}.



\subsection{GJ581 Activity}\label{subsec:GJ581}

GJ~581 is an M3V dwarf \citep{GJ581-spectraltype} at a distance of $6.2$~pc \citep{GJ581-distance} with mass $0.3 M_{\odot}$~\citep{GJ581-mass}, $\log g = 4.92$, and $T_{\rm eff} = 3480$~K \citep{GJ581-logg}. 
The number and locations of the planets orbiting GJ~581 have been extensively debated
~\citep{GJ581-Bonfils, GJ581-Udry,GJ581-Mayor,GJ581-Vogt,GJ581-Anglada,GJ581-Gregory,GJ581-Robertson}.
Confirmed planets are the Neptune analogs GJ~581~b ($P = 5.36$~days) and GJ~581~c ($P = 12.92$~days) and the potentially rocky super-Earth GJ~581~e at ($P = 3.15$~days) \citep{GJ581_carmenes}. \citet{GJ581-Robertson} suggested that the star is a differential rotator, but \citet{sdr} used Welch's power spectrum estimates and magnitude-squared coherence between H$\alpha$ and RV to argue that the star's single rotation period is $132$ days. If so, the disputed planets d and g are actually rotation harmonics. Here we calculate red-noise FALs for the GLS periodogram of the H$\alpha$ index measured from by \citet{GJ581-Robertson} (Figure \ref{fig:newts}).


\begin{figure}[h]
    \centering

     \includegraphics[width=\linewidth]{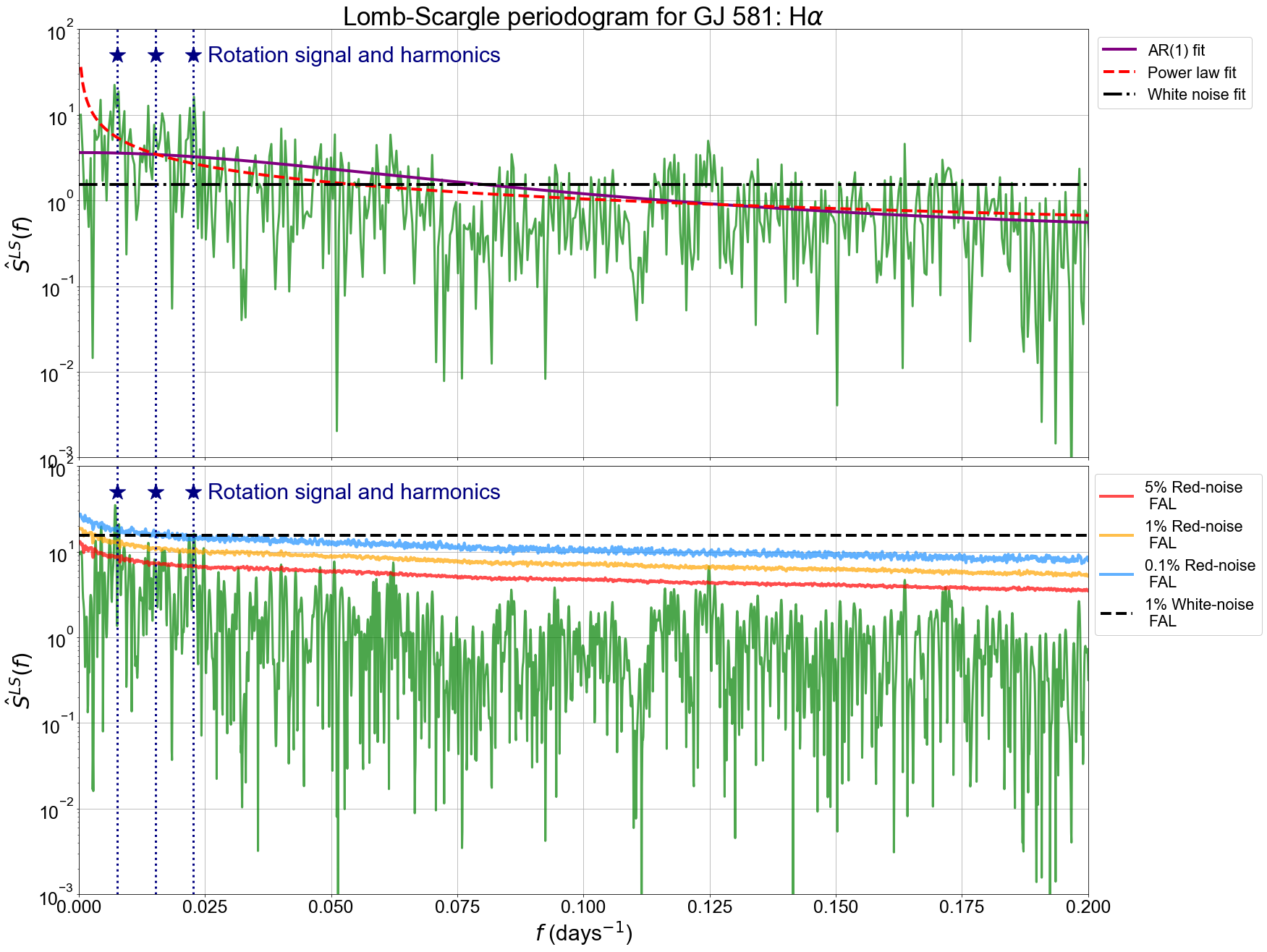}    
     \caption{GJ~581 H${\alpha}$: {\bf Top} --- The GLS periodogram $\hat{S}^{LS}(f_k)$~(green) with the best-fit AR(1)~(purple solid line), power law~(red dashed line) and white noise (black dash-dot line) models. Navy dotted lines mark the rotation signal and its first and second harmonics. {\bf Bottom} --- The red-noise $5\%$~(red), $1\%$~(orange), $0.1\%$~(blue) FALs, and the $1\%$ white-noise FAL (black dashed line) overlaid on the GLS periodogram. The rotation signal reaches a $99.9\%$ significance level according to both the red- and white-noise FALs. 
    }
    \label{fig:GJ581Ha}
\end{figure}

The top panel of Figure~\ref{fig:GJ581Ha} shows the H$\alpha$ GLS periodogram and the best-fit AR(1), power law, and white noise models. Since the periodogram has a lower dynamic range than those we have examined so far, the AR(1) and power law have the lowest best-fit values of $\hat{\phi}$ and $\hat{p}$ of any star in Table \ref{tab:tab1}. This ``whitening'' may be physically real, or it may come from the fact that GJ~581 H$\alpha$ has the largest ratio of RMS error bar to time series standard deviation of any dataset analyzed here. For example, $\alpha$~Cen~B $\log R^{^\prime}_{HK}$ has $\sqrt{(\frac{1}{N}\sum_{n=0}^{N-1} \sigma_n^2}) \bigg/ \left(\frac{1}{N-1} \sqrt{\sum_{n=0}^{N-1} (x_n - \bar{x})^2} \right) = 0.0083$ (where $\sigma_n$ is the error bar on the $n^{th}$ observation) while for GJ~581 H$\alpha$ the same ratio is 0.92. Large measurement uncertainties can mask variations caused by a physical red noise process.
Figure~\ref{fig:wnll} (4th row) shows that the three $-\mathcal{L}(\hat{\theta})$ distributions calculated according to \S \ref{subsec:assessing} have significant overlap. We calculate FALs using the power law model, for which the $-\mathcal{L}(\hat{\theta})$ distribution has the lowest mean value.

The bottom panel of Figure~\ref{fig:GJ581Ha} shows the $5\%$~(red), $1\%$~(orange), and $0.1\%$~(blue) red-noise FALs overlaid on the GLS periodogram along with the bootstrap 1\% white-noise FAL. The rotation signal ($f_{\rm rot} = 1/132$~days) and its second harmonic ($f = 3/132$~days) are significant against the $0.1\%$ FAL according to both the red and white noise models, while the compound periodogram peak centered at the first rotation harmonic ($f = 2/132$~days) reaches the 1\% red-noise FAL. The complex shape of periodogram peaks is due to the 1-year alias. There is a candidate stellar signal at $f = 0.125$~day$^{-1}$ ($P = 8$~days) that also reaches the 1\% red-noise FAL. The inventory of low-frequency signal detections is not significantly affected by moving from white noise to red noise-based FALs, but the white noise model misses the possible 8-day signal.



\subsection{KIC 6102338 Flux}\label{subsec:KIC}

\begin{figure}[h]
    \centering
    
    \includegraphics[width=0.9\linewidth]{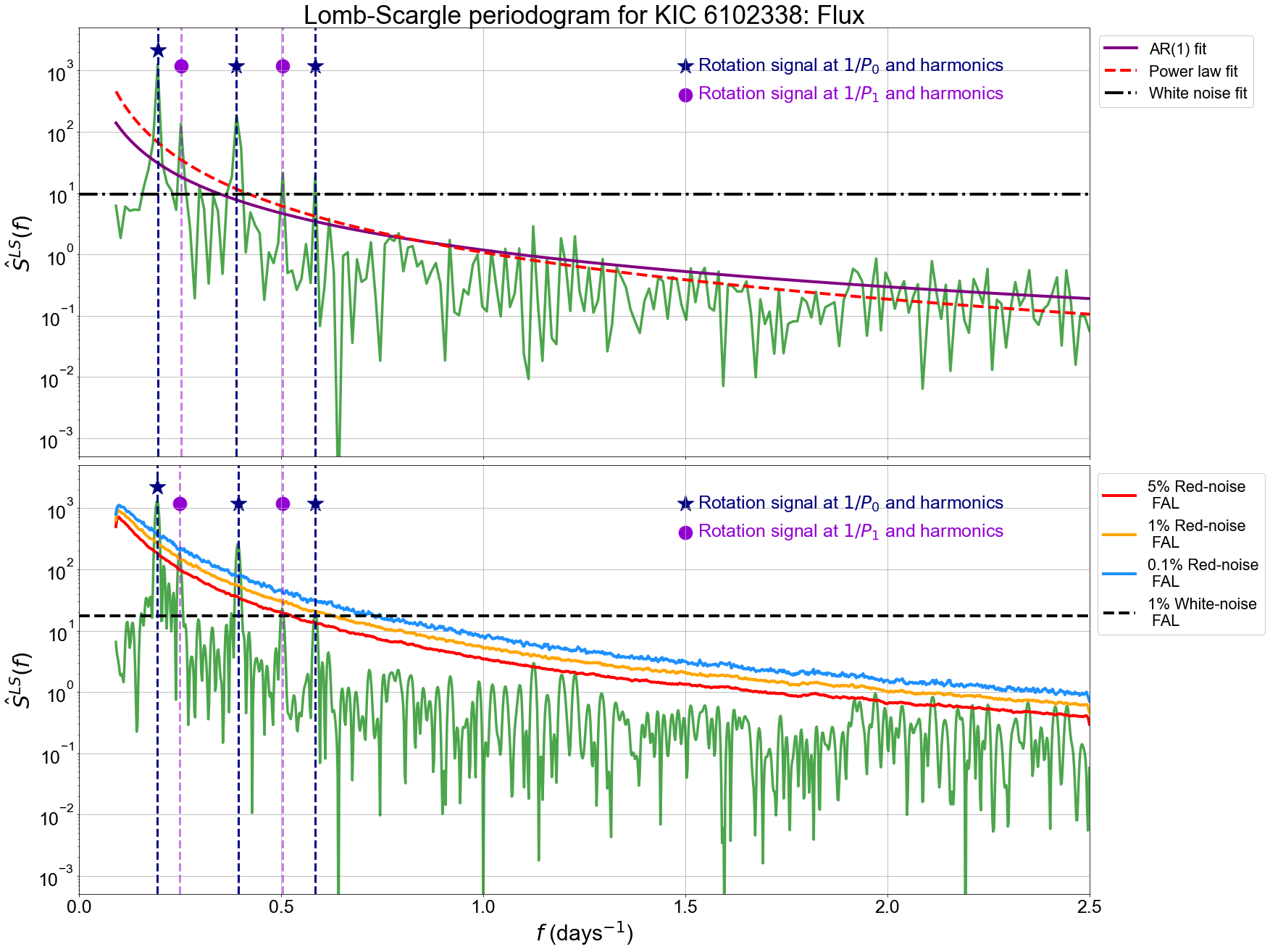}     
   
    \caption{{\bf Top} --- The GLS periodogram $\hat{S}^{LS}(f_k)$~(green) of KIC~6102338 flux with the best-fit AR(1)~(purple solid line), power law~(red dashed line) and white noise (black dash-dot line) models. The star's rotation signals and harmonics with frequency~$1/P_{0}$ are represented by the vertical navy dashed lines and $1/P_{1}$ by vertical violet dashed lines, {\bf Bottom} --- The red-noise $5\%$~(red), $1\%$~(orange), $0.1\%$~(blue) FALs and the $1\%$ white-noise FAL (black dashed line) are shown. Peaks at frequency $1/P_{0}$ and its first harmonic $2/P_0$ reach the $99.9\%$ significance level while the signal at frequency~$1/P_{1}$ reaches the $99\%$ significance level in the red-noise FALs.}
    \label{fig:KIC}
\end{figure}

The next red-noise FAL example uses the {\it Kepler} Q3 observations of KIC~6102338 \citep{borucki10}, an M dwarf at $307$~pc \citep{KIC-data}. Using the same dataset, \citet{reinhold2013rotation} detected differential rotation with periods $P_{0}=5.27$~days and $P_{1}=4.07$~days.
Applying the harmonic F test to a multitaper power spectrum estimate  \citep{thomson82} calculated with the \citet{bronez1988spectral} and \citet{chave19} technique for time series with missing data, \citet{dodsonrobinson24} identified statistically significant signals at frequencies ($1/P_{0}$, $2/P_{0}$, $3/P_{0}$, $1/P_{1}$ and $2/P_{1}$)~day$^{-1}$. Although the {\it Kepler} time series has a well-defined Nyquist frequency of $f_{\rm Nyq} = 24$~day$^{-1}$, the GLS periodogram is white and featureless at high frequencies. Here we compute red-noise FALS at
frequencies $0.018$~day$^{-1}$--$2.5$~day$^{-1}$, which includes contributions from rotation and possibly granulation/supergranulation.

The top panel of Figure~\ref{fig:KIC} shows the GLS periodogram with the best-fit AR(1), power law, and white noise models.
The best-fit parameters $\hat{\theta}$ and the $-\mathcal{L}(\hat{\theta})$ values are listed in Table~\ref{tab:tab1}. 
Figure~\ref{fig:wnll}~(5th row) shows the best-fit $-\mathcal{L}(\hat{\theta})$ distributions created from Monte Carlo realizations of the time series, with the white-noise $-\mathcal{L}(\hat{\theta})$ in the inset plot. 
The bottom panel of Figure~\ref{fig:KIC} shows the $5\%$~(red), $1\%$~(orange), and $0.1\%$~(blue) FALs based on the power law model, which produces the lowest values of $-\mathcal{L}(\hat{\theta})$. The $0.1\%$ white-noise FAL~(black dashed line) is also shown for comparison. The rotation signals at frequencies $1/P_{0}$ and $2/P_0$ reach the 0.1\% red-noise FAL ($99.9\%$ significance level), while the signals at $1/P_1$ and $3/P_0$ reach the 1\% FAL (99\% significance level). According to the white noise model, the entire continuum from $0.14 < f < 0.24$~day$^{-1}$ is statistically significant at the 1\% level. However, the power law model gives us the ability to prewhiten the power spectrum estimate and pick out only the statistically significant peaks, which are distinct from the continuum background and correspond to known oscillatory processes.


\subsection{HD 192310 radial velocities}\label{subsec:HD192310}


\begin{figure}[h]
    \centering
    \includegraphics[width=\linewidth]{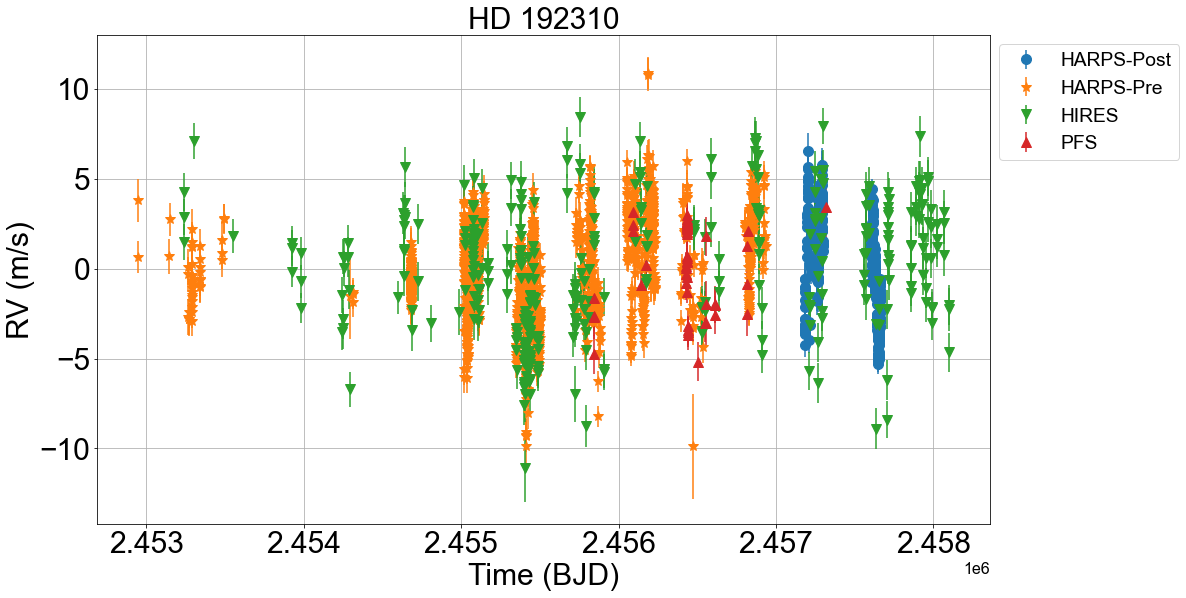}
    \caption{HD 192310 radial velocities from HARPS pre-2004 upgrade (HARPS-Pre: orange stars), HARPS post-2004 upgrade (HARPS-Post: blue dots), HIRES (green triangles), and PFS (red triangles). RV data are from \citep{Laliotisetal2023AJ....165..176L}.}
    \label{fig:HD192310}
\end{figure}


HD 192310 (Gl 785), a K3V type star of age 9.3 Gyr, mass $0.84~M_{\odot}$, $\log g = 4.5$, and $T_{\rm eff} = 5181$~K \citep{1982mcts.book.....H, 2007ApJS..168..297T, 2015Turnbull, Rosenthal2021}, is located at a distance of $8.8$~pc \citep{GJ581-distance}. 
\citet{Howard2011} announced the discovery of HD 192310 b, a Neptune-mass planet with a period of 74 days that was subsequently confirmed by \citet{Pepe2011} and \citet{Rosenthal2021}. \citet{Pepe2011} also reported another long-period planet, HD 192310 c, with a period of 525.8 days. Here we analyze HD 192310 radial velocities from three different instruments: (i) HARPS~\citep{2003Mayor}, (ii) HIRES~\citep{Vogt1994}, and (iii) Planet Finder Spectrometer (PFS)~\citep{2006Crane,2008Crane,2010Crane} presented by \citet{Laliotisetal2023AJ....165..176L}. RVs are plotted in (Figure~\ref{fig:HD192310}). We do not use the UCLES \citep[University College London Echelle Spectrograph;][]{1990SPIE.1235..562D} data from~\citet{Laliotisetal2023AJ....165..176L} due to the larger mean squared error bar ($\sim 1.8$~m~s$^{-1}$, compared with $\sim 0.8$~m~s$^{-1}$ for HARPS).

\begin{figure}[h]
    \centering
    
    \includegraphics[width=\linewidth]{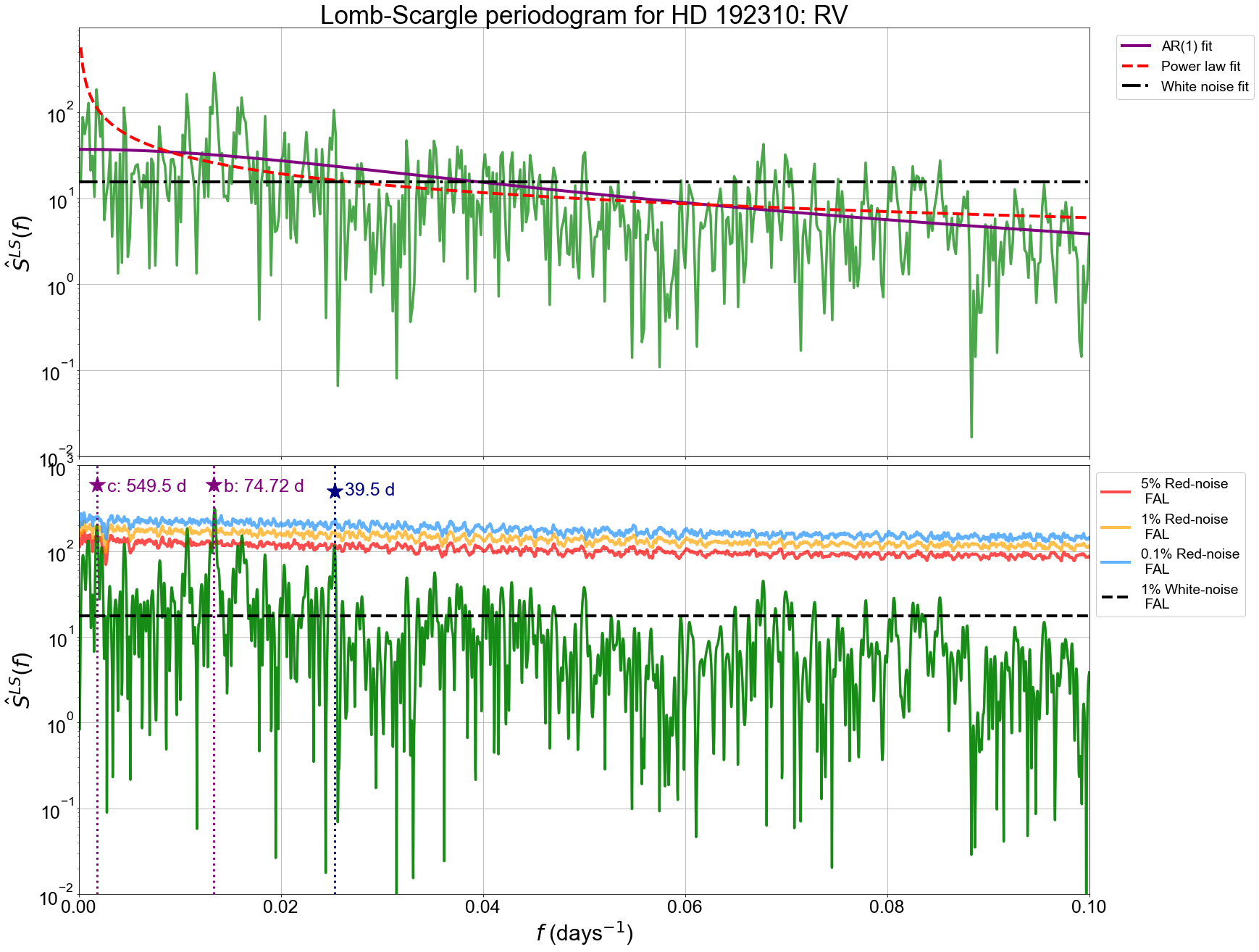}     
   
    \caption{{\bf Top} --- The GLS periodogram $\hat{S}^{LS}(f_k)$~(green) of HD 192310 radial velocities with the best-fit AR(1)~(purple solid line), power law~(red dashed line) and white noise (black dash-dot line) models. {\bf Bottom} --- The red-noise $5\%$~(red), $1\%$~(orange), $0.1\%$~(blue) FALs and the $1\%$ white-noise FAL (black dashed line) are shown. The peak associated with HD 192310 b (purple dotted lines) is $99.9\%$ significant, while peak with period of HD 192310 c (purple dotted lines) reaches the $99\%$ significance level. The differential rotation of the star is $95\%$ against the red noise FALs with a period of 39.5 d (navy dotted line).}
    \label{fig:HD192310_fals}
\end{figure}


The top panel of Figure~\ref{fig:HD192310_fals} shows the GLS periodogram $\hat{S}^{LS}(f_k)$ (green) of the combined RV data shown in Figure~\ref{fig:HD192310}, with its best-fit AR(1) (purple line), power law (red dashed line) and white noise (black dash-dot line) models. The best-fit model parameters $\hat{\theta}$ and the Whittle NLL $-\mathcal{L}(\hat{\theta})$ for all three models are presented in Table~\ref{tab:tab1}. Figure~\ref{fig:wnll} (bottom panel) shows that the AR(1) is the best-fit model, yielding the lowest values of $-\mathcal{L}(\hat{\theta})$, followed by the power law and white noise model.



The bottom panel of Figure~\ref{fig:HD192310_fals} shows the $5\%$~(red), $1\%$~(orange), and $0.1\%$~(blue) FALs based on the AR(1) model. The peak associated with the period of planet b reaches the 0.1\% red-noise FAL ($99.9\%$ significance level), while planet c is also detected at with $99\%$ significance at a period of 549.5~d as reported in \citet{Laliotisetal2023AJ....165..176L}. Other than the planet signals, the one-year aliases surrounding the planet b peak at 74.72~d also exceed the $5\%$ red-noise FAL. These alias peaks come from the same pseudowindow seen in the RV challenge dataset (\S \ref{subsec:rvchallenge}). We also detect a 39.5 d signal against the $5\%$ red-noise FAL. \citet{Laliotisetal2023AJ....165..176L} reported the same signal, which they attributed to differential rotation.

For the HD~192310 GLS periodogram, the bootstrap 1\% white-noise FAL not only does not match the continuum slope; it also has an unrealistic normalization. According to \citet[Chapter 9]{chernick08}, correlated noise is a known failure mode that requires bootstrap estimates of population statistics to be re-normalized.
The periodogram has $> 15$ peaks plus wide continuum regions that appear to be significant against the 1\% white-noise FAL, a result that has no reasonable physical interpretation. Our red-noise FALs account for the slope of the GLS periodogram continuum while also accurately representing the total signal power.

Another important point is that Figure \ref{fig:HD192310_fals} shows the periodogram of the RVs before any planet or stellar signals are removed. The differential rotation and the two confirmed planet signals simultaneously exceed the red noise 5\% FAL ($>95\%$ significance threshold) without requiring iterative model fitting and subtraction. Since an error in an early model-fitting step compromises all subsequent analysis, methods that facilitate identifying multiple significant periodicities in a single power spectrum estimate should be encouraged.


\section{Limitations and comparison with other methods}\label{sec:comparison}

Here we compare our prewhitening-based method with existing techniques for detecting periodic signals in time series with red noise. In RV planet searches, red noise
is typically treated by either modeling the time series' covariance matrix 
\citep[e.g.][]{baluev2011, Haywood2014} or using Monte Carlo methods to estimate parameters of stationary random processes in the time domain \cite[e.g.][]{tuomi12, TuomiAnglada_2013GJ163}. A partial list of references on time-domain red noise modeling with autoregressive, moving average, and Gaussian process models includes \citet{Toumi2013_SignalstauCeti, TuomiJenkins2014_COROT, Feroz2014, JenkinsTuomi_2014, Feng, delisle2, delisle3, Bortle_2021, suarez}. On the transit side, one red noise treatment comes from \citet{Csizmadia2022_I}, who modeled TESS photometry of HD~31221 by defining a likelihood function in a wavelet basis with an assumed power-law continuum PSD \citep{Carter_2009}.

The red noise treatment that is most similar to ours
comes from \citet{delisle1}. Following \citet{Baluev2008}, the observer constructs two models: base model $\mathcal{H}$,
which treats instrumental offsets and long-term drift, and enhanced model $\mathcal{K}(f)$ consisting of the base model plus the sinusoid $A \cos(2 \pi f t_n) + B \sin(2 \pi f t_n)$. 
In the \citet{Baluev2008} method, the best-fit parameters of $\mathcal{H}$ and $\mathcal{K}(f_k)$ are found via ordinary least-squares fitting (OLS). The periodogram is defined such that $z(f_k) \propto (\chi^2_\mathcal{H} - \chi^2_{\mathcal{K}}(f_k)) / \chi^2_{\mathcal{H}}$, where $\chi^2_\mathcal{H}$ is the $\chi^2$ of the residuals after subtracting the best-fit base model and $\chi^2_{K}$ is the same for the enhanced model. An analytical expression for the FAP comes from applying extreme value theory to white noise.

To allow for correlated noise, \citet{delisle1} execute a change of variable 
using the Cholesky factorization of covariance matrix $C = L L^{\rm T}$. Correlated noise is encoded in the off-diagonal terms of $C$ and best-fit parameters for models $L^{-1} \mathcal{H}$ and $L^{-1} \mathcal{K}(f)$ come from OLS. False alarm probabilities are then calculated analytically as in \citet{Baluev2008}. For time series with uniform observing cadence, observers could use the \citet{Baluev2008} method to fit an AR(1) model: the associated column vector in $\mathcal{H}$ would take the simple form $\{ 0, x_0, x_1, x_2, \ldots, x_{N-2}\}^{\rm T}$ and $\hat{\phi}$ would be its coefficient in the best-fit parameter vector. But when the time series has nonuniform observing cadence, the AR(1) model
is nonlinear in $\phi$ (Equation \ref{eq:ar1_ts}), which means $\phi$ cannot be estimated using either the \citet{Baluev2008} or \citet{delisle1} treatment and must be prescribed.

\citet{delisle1} used a Monte Carlo technique to test how incorrect correlated noise models affect the periodogram FALs by creating realizations of time series with covariance matrices of the general form
\begin{equation}
        C_{m,n} = \delta_{m,n}(\sigma_{n}^{2} + \sigma_{\rm jit}^{2}) + \sigma_{w}^{2}\exp\Bigg(-\dfrac{|t_{n}-t_{m}|}{\tau} \Bigg).
        \label{eq:cov}
\end{equation}
In Equation \eqref{eq:cov}, $\sigma_n^2$ and $\sigma_{\rm jit}^2$ represent instrumental errors and stellar jitter, respectively,\footnote{Time series and jitter units are arbitrary.} and $\sigma_w$ and $\tau$ are the white noise amplitude and persistence timescale.
Four different types of noise models were generated by changing covariance matrix parameters: 
\begin{enumerate}     
    \item Instrumental error only: $\sigma_{\rm jit} = \sigma_w = 0$
    \item Instrumental error and jitter: $\sigma_w = 0$
    \item Instrumental error and short-timescale correlated noise: $\sigma_{\rm jit} = 0$, $\sigma_w = 1$ and $\tau = 1$~days
    \item Instrumental error and long-timescale correlated noise: $\sigma_{\rm jit} = 0$, $\sigma_w = 1$ and $\tau = 30$~days
\end{enumerate}
For each noise model, 
$10^{6}$ realizations of $x_n$ were simulated using timestamps from the \citet{udry19} HARPS observations of HD~136352. FALs were estimated for {\it mismatched} $x_n$ and $C_{m,n}$ (for example, $x_n$ simulated according to $C_{m,n}$ type 3 but FALs calculated assuming $C_{m,n}$ type 4). 
\citet{delisle1} found that 
the detectability of long-period signals is reduced when $\tau$ is overestimated. On the other hand, underestimation of $\tau$
increases the false positive rate. 


Using our Whittle NLL-based prewhitening technique, the observer can numerically estimate nonlinear model parameters instead of prescribing them. Our method is therefore less likely to produce biased FALs than the \citet{delisle1} technique.
There is still the danger of choosing an incorrect model type (as opposed to simply assuming the wrong model parameters). But although we demonstrate the Whittle NLL using two simple power spectrum models, our method is general enough to fit any parametric model. The observer could also use the Whittle NLL to estimate correlated noise model parameters, then compute periodograms and FALs using the \citet{delisle1} method.

The observer must be mindful that the simple periodogram is a biased power spectrum estimator \citep{ThomsonandHaley}, so the estimated model parameters are also biased. 
For data with uniform observing cadence, bias can be substantially reduced by tapering~\citep[e.g.][]{Harris1978} and/or the Whittle NLL approximation can be debiased \citep{sykulski20}. Unfortunately it is not straightforward to generalize either operation to time series with uneven observing cadence. GLS periodograms inherit the bias problems that accompany simple periodograms and are also susceptible to
complex spectral window artifacts \citep{scargle1982, maxmoerbeck14, vanderplas}, all of which reduce the dynamic range of the power spectrum estimator \citep{uttley02, dodsonrobinson24}. 
One benefit of using our method is that the FALs have the same bias properties as the GLS periodogram. 
Even if the physical meaning of the spectral continuum is difficult to interpret, we can still 
estimate the power required for a periodic signal to exceed the continuum at a particular significance level. In Appendix \ref{app:quantiles}, we evaluate the FAL bias using quantiles of the $\chi^2_2$ distribution and discuss the connection between bias and error bars.

\section{Conclusion}\label{sec:conclusion}

This paper introduces a new method for estimating false alarm levels in Lomb-Scargle periodograms of time series with with correlated noise. 
In the statistics literature, prewhitening is the typical frequency-domain treatment of correlated noise: the observer subtracts the best-fit continuum model from $\log [\hat{S}(f)]$ and searches for peaks in the residual power spectrum that exceed significant $\chi^2$ values. Our method is a variant of prewhitening in which we fit the power spectrum continuum but do not subtract the best-fit model from the log-periodogram; instead, we use a Monte Carlo technique to estimate frequency-dependent FALs that have the same bias properties as the periodogram. 
While this paper is primarily directed at planet searches and studies of stellar activity, our method can be applied to any type of time series with uneven observing cadence. It can also be used to the estimate persistence time $\tau$ for a covariance-based red noise modeling technique such as that of \citet{delisle1}.


Our software offers the user a choice of two red noise models: i) AR(1) (autoregressive model of order 1), a persistence model in which each observation is partially predicted by the previous observation \citep{tauest,redfit,redfitx,Robinson,Hasslemann, Gilman, SmithAllen} and ii) a power law with a negative exponent, which is motivated by physical processes in the stellar photosphere~\citep{Frisch_1995, Kolmogorov,kippenhahn2012stellar, Cegla2018, Cegla2019, Paxton,turbulence,turbulence2,Cranmer}. 
To fit the noise models, we use the Whittle approximation to the negative log-likelihood \citep[Equation \ref{eq:wnll};][]{whittle,whittle_2,whittle1957curve,Whittle1953TheAO}, which has been widely applied to power spectrum analysis of uniformly spaced time series data in a variety of fields. This work is the first application of the Whittle NLL to unevenly spaced astronomical time series. Our method can be extended to fit more sophisticated models, such as ARMA(p,q), ARFIMA(p,q,d), or Matern.

Our red noise-based FALs perform significantly better than white noise-based approximations for time series with strong red noise backgrounds, such as the $\alpha$~Cen~B $\log R_{HK}^{\prime}$ observations of \citet{dumusque}, System 11 from the RV fitting challenge \citep{RVchallengedata}, and {\it Kepler} observations of the rapid rotator KIC 6102338 \citep{borucki10, reinhold2013rotation} (See Figure \ref{fig:wnll}). Our method recovers high-frequency signals (\S \ref{subsec:acenb}) while rejecting the low-frequency continuum that is significant against the white noise FALs. 
In contrast, the Lomb-Scargle periodogram of the \citet{GJ581-Robertson} GJ~581 H${\alpha}$ time series is not significantly red, as we see from the overlapping Whittle NLL distributions from AR(1), power law, and white noise model fits (Figure~\ref{fig:wnll}). Since AR(1) (with $\phi=0$) and power law models (with $p=0$) can both represent white noise, our method can calculate FALs for periodograms of time series with either correlated or uncorrelated noise processes.
For HD 192310 RVs (\S \ref{subsec:HD192310}), the red-noise FALs identify the planet and differential rotation signals reported by \citet{Pepe2011}, \citet{Howard2011}, and \citet{Laliotisetal2023AJ....165..176L} at a 95\% significance level without requiring iterative model fitting and subtraction (Figure \ref{fig:HD192310_fals}, bottom panel). 
We encourage astronomers to use Whittle NLL-based FALs that are general enough to treat red noise instead of white noise-based FALs to identify significant signals in Lomb-Scargle periodograms.

\section*{Software Availability} 
The calculations featured in this paper are done using the python package RedNoiseFALs.py (https://doi.org/10.5281/zenodo.15881590). This GitHub repository contains a demo Jupyter Notebook that demonstrates the software functionality and notebooks that perform the calculations included in this article. 
\software{Astropy \citep{astropy:2013, astropy:2018, astropy:2022}}.

\section*{Data Availability}
The Kepler observation used in section~\ref{subsec:KIC} are archived at (doi:10.17909/0etb-4t73). The $\alpha$ Cen B data that was used in sections \ref{sec:mm} and \ref{eqn:acenbsin} can be found in the supplementary information section of \citet{dumusque}. The RV challenge data set used in section \ref{subsec:rvchallenge} can be accessed at VizieR DOI~\citet{vizier:J/A+A/593/A5}. The GJ 581 H$\alpha$ data set used in section \ref{subsec:GJ581} can be found in the supplementary materials of \citet{GJ581-Robertson}. The HD~192310 RV data is available at~\citet{laliotis-data}.

\section*{Acknowledgment}
This research has made use of the VizieR catalogue access tool, CDS, Strasbourg, France (DOI : 10.26093/cds/vizier). The original description of the VizieR service was published in 2000, A\&AS 143, 23. 

\facilities{ESO:3.6m (European Southern Observatory (ESO) 3.6m Telescope at La Silla Observatory), HET (University of Texas, Austin 9.3m Hobby-Eberly Telescope at McDonald Observatory), Kepler (NASA 0.95m Kepler Satellite Mission)}

This work is supported by the National Science Foundation (NSF) grant 2307978.

This material is based on work supported by the U.S. Department of Energy, Office of Science, under contract number DE-AC02-06CH11357.


\appendix

\section{Assessing the effects of error bars using $\chi^2$ quantiles}
\label{app:quantiles}

In \S \ref{subsec:calculating} we discussed how error bars tend to ``whiten'' the Monte Carlo-derived FALs. Since the PSD-normalized Lomb-Scargle periodogram is $\chi^2_2$ distributed under the null hypothesis of white noise \citep{scargle1982}, such that $\hat{S}^{LS}(f) / S(f) \sim \chi^2_2 / 2$, the whitening effect can be measured using $\chi^2_2$ quantiles. 
Recall that our FAL-calculation method is a modified version of prewhitening in which we fit, but do not subtract, a model of the log power spectrum. If instead we proceed with model subtraction, a {\it perfect} model fit should yield a white residual log-periodogram with Monte Carlo FALs that match the $\chi^2_2$ quantiles.

\begin{figure}
    \centering
    \includegraphics[width=\textwidth]{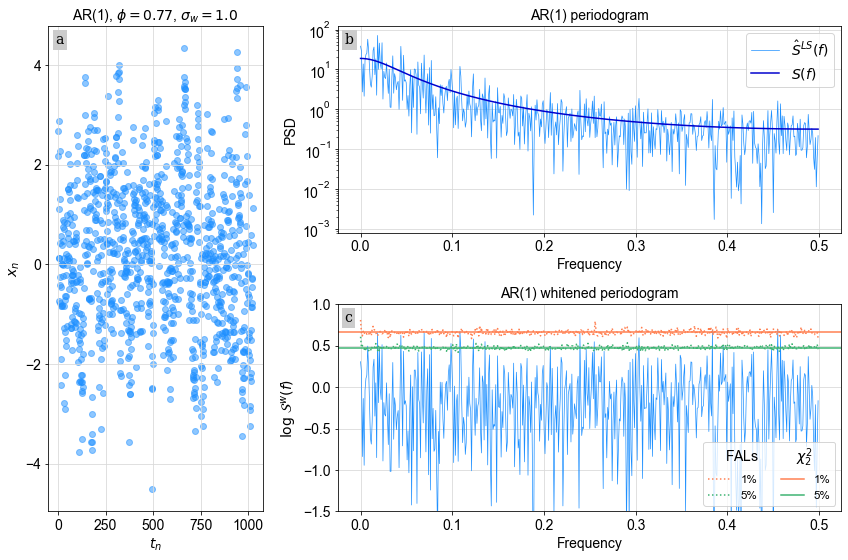}
    \caption{{\bf a:} Time-domain realization of an AR(1) process with $\phi = 0.77$ and $\sigma_w = 1.0$. {\bf b:} Periodogram $\hat{S}^{LS}(f)$ (light blue) and true power spectrum $S(f)$ (dark blue) of time series in panel a. {\bf c:} Whitened log-periodogram $\log \left[ \hat{S}^{LS}(f) \right] - \log \left[ S(f) \right]$ overlaid with 1\% and 5\% $\chi^2_2$ quantiles (orange and green solid lines, respectively) and 1\% and 5\% whitened Monte Carlo FALs (orange and green dotted lines).}
    \label{fig:AR1_prewhitening}
\end{figure}

Figure \ref{fig:AR1_prewhitening} shows the start-to-finish prewhitening procedure for a synthetic time series created from an AR(1) process with $\phi = 0.77$ and $\sigma_w = 1.0$ sampled with uniform timesteps and no errors.\footnote{Realizations of AR(1) processes on uniform time grids are generated in \texttt{python} with \texttt{statsmodels.tsa.arima\_process.ArmaProcess}.} The time series $x_n$ (panel a) has true power spectrum $S(f)$ (panel b, dark blue) and generates periodogram $\hat{S}^{LS}(f)$ (panel b, light blue). (All experiments in this section use the same $x_n$.) Panel c shows the whitened log-periodogram $\log \mathcal{S}^w(f) = \log \left[ \hat{S}^{LS}(f) \right] - \log \left[ S(f) \right]$ (light blue) plus horizontal lines at the theoretical 1\% and 5\% FALs (orange and green, respectively). The theoretical FALs are $\log \left( \chi^2_2/2 \right)$ values corresponding to $1-\operatorname{CDF}(\chi^2_2) = \alpha$, where CDF is the cumulative distribution function and $\alpha$ is the risk level (1\% or 5\%). From an ensemble of periodograms of 1000 synthetic time series drawn from the same the AR(1) process (i.e.\ Equations \ref{eq:ar1_ts}--3 with $\phi = 0.77$ and $\sigma_w = 1.0$ are applied 1000 times), we compute Monte Carlo FALs by finding the empirical 99\% and 95\% quantiles at each frequency $f_k$. The dotted lines in Figure \ref{fig:AR1_prewhitening}c show the whitened Monte Carlo log-FALs, $\log[\operatorname{FAL}] - \log [S(f)]$, again with 1\% in orange and 5\% in green. The agreement between the theoretical and Monte Carlo FALs is excellent.

Now suppose samples from the same AR(1) process have small, uniform error bars $\sigma_n = 0.5$. We create a synthetic dataset $x^{\prime}_n = x_n + \mathcal{N}(0, 0.5)$ which has periodogram $\hat{S}^{\prime LS}(f)$, then calculate frequency-dependent 1\% and 5\% FALs from the periodograms of 1000 more realizations of $x^{\prime}_n$. To whiten $\log \left [\hat{S}^{ \prime LS}(f) \right ]$ and its log-FALS, we proceed as if we do not know the true power spectrum of the underlying AR(1) process and find the best-fit model parameters using our \texttt{RedNoiseFALs} software.

\begin{figure}
    \centering
    \includegraphics[width=\textwidth]{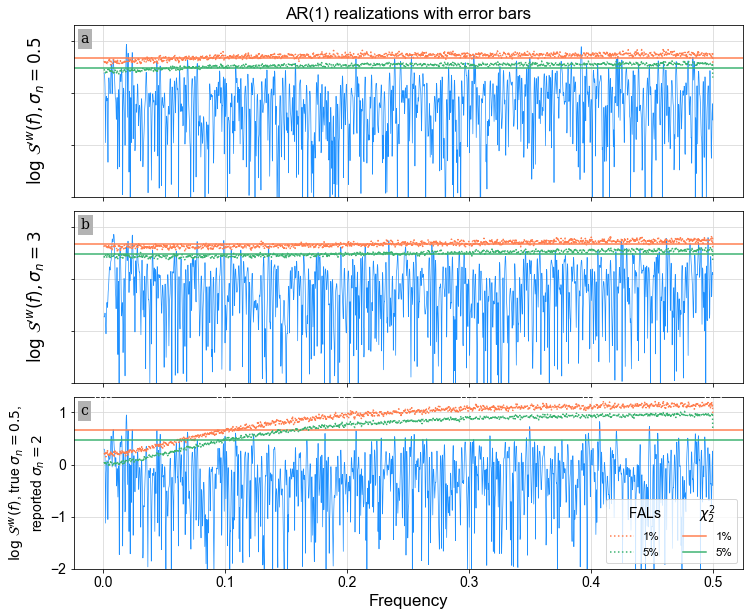}
    \caption{{\bf a:} Whitened log-periodogram $\log \mathcal{S}^{\prime w}(f)$ and FALs for a time series drawn from the AR(1) process shown in Figure \ref{fig:AR1_prewhitening} ($\phi = 0.77, \sigma_w = 1$). Error bars are $\sigma_n = 0.5$. {\bf b:} Same as above but with $\sigma_n = 3$. {\bf c:} Whitened log-periodogram and FALs for a time series for which the error bars are overestimated. The dataset $x^{\prime}_n$ has true $\sigma_n = 0.5$, but FALs are calculated using $\sigma_n = 2$.}
    \label{fig:AR1_prewhitening_errors}
\end{figure}

Figure \ref{fig:AR1_prewhitening_errors}a shows the whitened log periodogram $\log \mathcal{S}^{\prime w}$ (blue solid line), whitened Monte Carlo FALs computed as in \S \ref{subsec:calculating} (dotted lines), and theoretical $\chi^2_2$ FALs (solid lines). Here there is a modest upward trend in the log-FALs after the best-fit $\log [S(f)]$ is subtracted, indicating that the frequency-dependent FALs produced by our software are ``whiter'' than the AR(1) model would predict (see \S \ref{subsec:calculating}). The effect on oscillation detection is that a high-frequency signal must have a higher peak-to-continuum ratio than a low-frequency signal in order to exceed a FAL and be considered statistically significant. Such FALs are accurate reflections of observational uncertainty: when the process under observation has a red power spectrum but the error bars are (assumed to be) Gaussian, the periodogram signal-to-noise ratio monotonically decreases with frequency, complicating detection of high-frequency signals.


Repeating the same experiment with larger error bars, $\sigma_n = 3$, gives the whitened log-periodogram shown in Figure \ref{fig:AR1_prewhitening_errors}b. We see the same effect as in Figure \ref{fig:AR1_prewhitening_errors}a, a small upward slope in Monte Carlo $\log[\operatorname{FAL}] - \log [S(f)]$, indicating that the FALs are whiter than the best-fit model would predict. Here again the slight mismatch between the log-FALs and their associated $\chi^2_2$ quantiles illustrates the effects of Gaussian measurement errors on observations of a process with a red power spectrum.

While underestimated error bars are always problematic, leading observers to overestimate detection significance in any context, there is a problem posed by {\it overestimated} error bars that is specific to red noise FALs. Figure \ref{fig:AR1_prewhitening_errors}c shows $\log \mathcal{S}^{\prime w}(f)$ from a realization of the same $x^{\prime}_n$, $\sigma_n = 0.5$ process examined in Figure \ref{fig:AR1_prewhitening_errors}a. However, we create the realizations of $x^{\prime}_n$ used to calculate FALs assuming $\sigma_n = 2$, a 16-fold overestimate of the instrumental noise variance. The ``whitened'' log-FALs do not come close to matching the $\chi^2_2$ quantiles: signal significance is badly underpredicted at high frequencies and overpredicted at low frequencies. Unlike panels a and b, the log-FALs in panel c do not approximate straight lines. From Figure \ref{fig:AR1_prewhitening_errors}c, we infer that overestimated error bars destroy the red noise FALs because they do not accurately reflect the signal-to-noise ratio of the periodogram. Overestimated error bars have the undesirable property of triggering false positives {\it and} false negatives. Observers who suspect error bar overestimation may wish to re-estimate error bars before using \texttt{RedNoiseFALs}.

\pagebreak




\bibliography{sample631}{}
\bibliographystyle{aasjournal}


\end{document}